\documentclass[useAMS,usenatbib]{mn2e}

\usepackage{graphicx}
\usepackage[dvips]{color}
\usepackage{epsfig}

\title[Optical and Infrared Diagnostics of Low Redshift SDSS galaxies in the SWIRE Survey]{Optical and Infrared Diagnostics of SDSS galaxies in the SWIRE Survey}
\author[Davoodi et al.]
{P. Davoodi$^{1}$\thanks{E-mail:
payam@sussex.ac.uk}, F. Pozzi$^{2}$, S. Oliver$^{1}$, M. Polletta$^{3}$, A. Afonso-Luis$^{4}$, D. Farrah$^{5}$,
\newauthor
 E. Hatziminaoglou$^{4}$, G. Rodighiero$^{6}$, S. Berta$^{6}$, I. Waddington$^{1}$, C. Lonsdale$^{3,7}$, 
\newauthor
M. Rowan-Robinson$^{8}$, D.L. Shupe$^{9}$, T. Evans$^{7}$, F. Fang$^{9}$, H.E. Smith$^{3}$, J. Surace$^{9}$\\
\\
$^{1}$Astronomy Centre, Department of Physics and Astronomy, University of Sussex, Brighton, BN1 9QH, UK\\
$^{2}$Dipartimento di Astronomia, Universit\`a di Bologna,
      viale Berti Pichat 6, I-40127 Bologna, Italy\\
$^{3}$Centre for Astrophysics \& Space Sciences, University of California, San Diego, La Jolla, CA 92093-0424, USA\\
$^{4}$Instituto de Astrof\'{\i}sica de Canarias,
        Via Lactea S/N, E-38200, La Laguna, Spain\\
$^{5}$Department of Astronomy, Cornell University, Ithaca, NY 14853, USA\\
$^{6}$Dipartimento di Astronomia, Universit\`a di Padova,
        Vicolo dell'Osservatorio 2, I-35122, Padova, Italy\\
$^{7}$Infrared Processing and Analysis Centre, California Institute of Technology, Pasadena, CA 91125, USA\\
$^{8}$Astrophysics Group, Blackett Laboratory, Imperial College London, Prince Consort Road, London, SW7 2BW, UK\\ 
$^{9}$$Spitzer$ Science Centre, California Institute of Technology, Pasadena, CA 91125, USA\\
}
  
\begin{document}

\date{Accepted . Received ; in original form }

\pagerange{\pageref{firstpage}--\pageref{lastpage}} \pubyear{2006}

\maketitle

\label{firstpage}

\begin{abstract}
We present the rest-frame optical and infrared colours of a complete sample of 1114 $z$$<$0.3 galaxies from the Spitzer Wide-area InfraRed Extragalactic Legacy Survey (SWIRE) and the Sloan Digital Sky Survey (SDSS).
We discuss the optical and infrared colours of our sample and analyse in detail the contribution of dusty star-forming galaxies and AGN to optically selected red sequence galaxies.

We propose that the optical (g-r) colour and infrared $\log$($L_{24}$/$L_{3.6}$) colour of galaxies in our sample are determined primarily by a bulge-to-disk ratio. The (g-r) colour is found to be sensitive to the bulge-to-disk ratio for disk-dominated galaxies, whereas the $\log$($L_{24}$/$L_{3.6}$) colour is more sensitive for bulge-dominated systems. 

We identify $\sim$18$\%$ (195 sources) of our sample as having red optical colours and infrared excess. Typically, the infrared luminosities of these galaxies are found to be at the high end of star-forming galaxies with blue optical colours. Using emission line diagnostic diagrams, 78 are found to have an AGN contribution, and 117 are identified as star-forming systems. The red (g-r) colour of the star-forming galaxies could be explained by extinction. However,
their high optical luminosities cannot. We conclude that they have a significant bulge component.  

The number densities of optically red star-forming galaxies are found to correspond to $\sim$13$\%$ of the total number density of our sample. In addition, these systems contribute $\sim$13$\%$ of the total optical luminosity density, and 28$\%$ of the total infrared luminosity density of our SWIRE/SDSS sample. These objects may reduce the need for ``dry-mergers".

\end{abstract}


\section{Introduction}

The global star-formation history of the universe can be investigated by combining ultraviolet and optical wavelength information with the infrared (IR), probing different stages (un-obscured and obscured) of star-formation processes.

The infrared satellite IRAS (InfraRed Astronomy Satallite - Neugebauer et al.\ 1984) has probed the local universe to determine how a third of the optical light from normal galaxies is absorbed by dust, and how this fraction can become higher when galaxies with the most active star-formation acitivity (LIGs, Sanders \& Mirabel,\ 1996) are considered. Both IRAS and the ISO satellite (Kessler et al.\ 1996) extended this analysis, showing the importance of dust in the high redshift  universe (Elbaz et al.\ 1999; Gruppioni et al.\ 2002; Rowan-Robinson et al.\ 2004; Metcalfe et al.\ 2005, Lonsdale et al.\ 2006).

Likewise, several studies (Wang \& Silk 1994, Hogg et al.\ 2004, McIntosh et al.\ 2005) have shown that optical colours alone are not sufficient to discriminate the nature of some galaxy samples.
In the K20 survey (Cimatti et al.\ 2004), spectral classifications show how two distinct populations with similar optical/near infrared colours contribute to a population of Extremely Red Objects (EROs) at high redshift ($z$$\sim$2): (i) old stellar systems with no signs of star-formation and (ii) dusty star-forming galaxies. Moreover, studies of dusty galaxies locally and at high redshift (Poggianti \& Wu 2000; Rigopoulou et al.\ 2000, Flores et al.\ 2004) have shown that more than 70$\%$ of the energy emitted by young stars is reprocessed in the infrared, leaving no trace in the optical spectrum of these systems. Given these results, complementing optical studies with infrared information has become a crucial part of improving our understanding of optically-selected galaxy samples. 

One of the main aims of this paper is to expand the infrared colour distribution and thus the star-formation activity and dust content of low redshift galaxies as a function of their rest-frame optical colours. Results from Strateva et al.\ (2001), Blanton et al.\ (2003), Bell et al.\ (2003b) and Baldry et al.\ (2004) indicate that the colour distribution of galaxies in the optical can be represented by a $bi-modal$ function. In other words, formation processes give rise to two dominant populations with different average colours and dispersions.  Galaxies with red optical colours, `red sequence', are generally considered to be early-type galaxies. Their colours are the result of old, high metallicity stellar populations with little ongoing star-formation. In comparison, galaxies with blue optical colours, `blue sequence', are considered to be late-type galaxies, probing current star-formation. This optical colour distribution is found to be bimodal at z$\leq$0.1 (Strateva et al.\ 2001; Baldry et al.\ 2004), z$\sim$1 (Bell et al.\ 2004) and at least out to z$\sim$2 (Giallongo et al.\ 2005), clearly reflecting the broad differences in star formation histories over large redshift ranges. 

Since most of the activity (e.g., star-formation and/or AGN emission) in galaxies is hidden by dust, the bolometric luminosity of active systems in the local universe and at high redshifts (e.g. Sanders et al.\ 2004) is mostly emitted in the infrared. Therefore, probing this emission will give a clearer insight into the star-formation activity in galaxies forming the two sequences, and allow us to investigate whether the luminosity density of local galaxies is underestimated in the optical.

To understand the nature of galaxies in the two sequences at low redshift, we study the infrared properties of galaxies from the Spitzer Wide-Area InfraRed Extragalactic survey (\renewcommand{\thefootnote}{\fnsymbol{footnote}}SWIRE\footnote[2]{\textit{www.ipac.caltech.edu/SWIRE/}} ; Lonsdale et al.\ 2003), associating sources with their optical counterparts from the Sloan Digital Sky Survey (SDSS, York et al.\ 2000). SWIRE, the largest of the six Spitzer Legacy programs, is a wide-area imaging survey, mapping the distribution of spheroids, disks, starbursts and active galactic nuclei (AGN) and their evolution from $z$$\sim$3 to the current epoch. The survey covers $\sim$49 square degrees (in 6 fields) in all seven Spitzer bands: 3.6, 4.5, 5.8, and 8$\mu$m with the InfraRed Array Camera ($IRAC$ - Fazio et al.\ 2004) and 24, 70 and 160$\mu$m with the Multi-band Imaging Photometer for Spitzer ($MIPS$ - Rieke et al.\ 2004), detecting $\sim$2.5 million galaxies down to $f_{3.6{\mu}m}$$\approx$ 5$\mu$Jy.

The large area of SWIRE provides statistically significant population samples over enough volume cells that we can resolve the star formation history as a function of epoch and environment. Combined with data from SDSS, the largest optical spectroscopic survey currently available, we can select a large sample of galaxies and efficiently probe their optical and infrared emission.

The paper is outlined as follows. $\S$2 describes the infrared and optical data used to construct our sample. Sections 3 and 4 discuss the techniques used to derive rest-frame luminosities and spectral classifications using diagnostic diagrams based on optical spectra. In Sections 5 and 6 we investigate the optical and infrared colours of our SWIRE/SDSS sample. In $\S$7 we analyse the properties of galaxies with red optical colours and high mid-infrared emission, to determine how they differ from star-forming galaxies with blue optical colours. The number density and optical/infrared luminosity density of the different populations in our sample are determined in $\S$8. $\S$9 presents discussions and conclusions.
For this work, we use a cosmological model with $\Omega$$_{0}$ = 0.3, $\Omega_{\Lambda}$ = 0.7, and a Hubble constant of H$_{0}$ = 72 km s$^{-1}$ Mpc$^{-1}$.

\section[]{The Data}

We use $IRAC$ (3.6$\mu$m, 4.5$\mu$m, 5.8$\mu$m, 8.0$\mu$m) and $MIPS$ (24$\mu$m, 70$\mu$m, 160$\mu$m) catalogues from the northern SWIRE fields of Lockman, ELAIS-N1 and ELAIS-N2, cross-correlated with SDSS $ugriz$ photometry and spectroscopic catalogues to provide a complete 13$\le$$r$$\le$17.5 sample of 1114 galaxies at z$<$0.3. The total survey area of the three northern SWIRE fields is $\sim$24 square degrees, of which the overlap with SDSS is $\sim$18 square degrees.

Here we discuss the technical aspects of the SWIRE and SDSS data sets.

\subsection{Lockman, ELAIS-N1 and ELAIS-N2}

The SWIRE Lockman field is centred at $10^{h}$$45^{m}$$00^{s}$
+$58^{d}$$00^{m}$$00^{s}$, with coverage of $\sim$10.6 square degrees. $IRAC$
(3.6$\mu$m, 4.5$\mu$m, 5.8$\mu$m and 8$\mu$m)+$MIPS$ (24$\mu$m) observations were performed on 2004 April and 2004 May.

The SWIRE ELAIS-N1 field is centred at $16^{h}$$11^{m}$$00^{s}$
+$55^{d}$$00^{m}$$00^{s}$, with coverage of $\sim$9 square degrees. $IRAC$
+24$\mu$m observations were performed in 2004 January and 2004 Feburary.

The SWIRE ELAIS-N2 field is centred at $16^{h}$$36^{m}$$48^{s}$
+$41^{d}$$01^{m}$$45^{s}$, with coverage of $\sim$4 square degrees. $IRAC$+24$\mu$m observations were performed in 2004 July.

The average 5$\sigma$ depths of Lockman, ELAIS-N1 and ELAIS-N2 are 5.0, 9.0, 43, 40 and
311$\mu$Jy at 3.6, 4.5, 5.8, 8, and 24$\mu$m respectively (Surace et al.\ 2005), consistent with the 90$\%$ completeness levels for source extraction. Fluxes were extracted in 5.8$\arcsec$ radius apertures for
$IRAC$ ($\sim$2x the FWHM beam) and 12$\arcsec$ for $MIPS$, using SExtractor 
(Bertin and Arnouts,\ 1996). The absolute flux calibrations are correct within $\sim$10$\%$ for $IRAC$ and $MIPS$ 24$\mu$m channel data, and were
confirmed for $IRAC$ and $MIPS$ 24$\mu$m by comparison to 2MASS. Further discussion on the
data processing is given by Surace et al.\ (2005, 2006) and Shupe et al.\ (2006). 

The Lockman field contains 681,587 SWIRE sources. The fields of ELAIS-N1 and ELAIS-N2 contain 
411,015 and 309,507 SWIRE sources. Therefore, our total SWIRE sample consists of $\sim$1.4 million sources surveyed over a total of  $\sim$24 square degrees.

The 5$\sigma$ depths of $MIPS$ 70$\mu$m and 160$\mu$m data are 14mJy and 105mJy in all three SWIRE fields, consistent with 60$\%$ completeness levels for source extraction (Afonso-Luis et al.\ 2006, in prep.). PRF (Point-source Response Function) fluxes in both bands were extracted using APEX. Further details on flux calibrations and source extraction can be found in Frayer et al.\ (2006) and (Afonso-Luis et al.\ 2006). The Lockman, ELAIS-N1 and ELAIS-N2 fields contain 2397, 1130 and 2485 sources with 70$\mu$m detections. With regards to 160$\mu$m detections, the fields of Lockman, ELAIS-N1 and ELAIS-N2 contain 1076, 439 and 1106 sources.

\subsection{SDSS data}

The SDSS sample of spectroscopically observed galaxies (Strauss et al. 2002) taken from the third Data Release (DR3; Abazajian et al. 2004) was used to provide optical $ugriz$ magnitudes and spectroscopic redshifts for our sample. The SDSS data provides full coverage of the SWIRE fields of Lockman and ELAIS-N2 but covers only a third of the ELAIS-N1 field. Since our initial analysis was carried out using DR3, we have continued to use DR3 since DR4 does not provide any additional coverage of the SWIRE fields.

To remove sources with unreliable redshifts from our optical sample, we eliminated sources with the following SDSS spectroscopic pipeline flags; ZWARNING$\neq$0 (problem with the redshift fitting) and  zConf $<$ 0.85 (low confidence in the spectroscopic redshift). To eliminate stars, sources were only considered if they had a spectroscopic object class specClass = GALAXY or QSO and spectroscopic redshift $>$ 0.001 (see Blanton et al.\ 2004). Approximately 5$\%$ of the SDSS sample did not meet this criteria, and were therefore classified as stars. These sources were removed from the SDSS sample.

\subsection{SDSS morphological and spectral classifiers}

The SDSS collaboration provide three galaxy classifiers: $C_{r}$ ($r$-band concentration index - Strateva et al.\ 2001, Shimasaku et al.\ 2001), S$\acute{{e}}$rsic index (Blanton et al.\ 2003c, Hogg et al.\ 2004) and eClass (Yip et al.\ 2004, Connolly $\&$ Szalay 1999 and Connolly et al.\ 1995). 

The $r$-band concentration index $C_{r}$ is a morphological classifier, defined as the ratio of the radii containing 90$\%$ and 50$\%$ of the Petrosian $r$-band galaxy light. 
Shimasaku et al.\ (2001) and Strateva et al.\ (2001) find reasonable correlation between qualitative morphological classifications and $C_{r}$, with a $C_{r}$$\ge$2.6 selection for early-type galaxies. 

S$\acute{{e}}$rsic index is another parameter measuring galaxy morphology, derived by fitting the functional form $I(r)$=$I_{0}$exp(-r$^{1/n}$) (where $n$ is the S$\acute{{e}}$rsic index itself), to each galaxy surface brightness profile. Hogg et al.\ (2004) find that galaxies with purely exponential profiles (S$\acute{{e}}$rsic index $n$$\sim$1) are considered late-type (i.e. disk dominated) while galaxies with a de Vaucouleurs profiles (S$\acute{{e}}$rsic index $n$$\sim$4) are considered early-type (i.e. bulge dominated).

The SDSS single parameter spectral classifier eClass is based on Principal Component Analysis, using cross-correlation with eigentemplates constructed from SDSS galaxy spectra. This parameter ranges from approximately $-$0.35 to 0.5 for early to late galaxies (Stoughton et al.\ 2004).  Based on the work by Yip et al.\ (2004) where eClass is modelled as a function of the distribution of galaxy populations, galaxies with an eClass $<$ $-$0.02 are considered to be early-type.

For the initial work discussed in $\S$5, we define galaxies with $C_{r}$$\ge$2.6 and eClass $<$ $-$0.02 as early-type, and those with $C_{r}$$<$2.6 and eClass $\ge$ $-$0.02 as late-type. However, there are also sources that do not satisfy both conditions. The S$\acute{{e}}$rsic will be used as an additional classifier, and will be discussed in $\S$6. 

\subsection{SWIRE/SDSS sample association}

SWIRE $IRAC$+24$\mu$m bandmerged catalogues for Lockman, ELAIS-N1 and ELAIS-N2 were initially matched with the SDSS catalogues. Using the $r$-band SDSS and the 3.6$\mu$m SWIRE co-ordinates, a SWIRE galaxy within 1.5\arcsec of the SDSS co-ordinate is chosen as the best match. This search radius is similar to that used to match SWIRE optical and infrared data sets (see Surace et al.\ 2005). Since our sample consists of bright sources, we have verified that such criteria are adequate for matching sources. 
The $MIPS$ 70$\mu$m and 160$\mu$m source catalogues were then individually matched to the 3.6$\mu$m SWIRE co-ordinates of the SWIRE/SDSS catalogue.
The $MIPS$ 70$\mu$m source catalogue was matched with search radius of 10\arcsec. Sources were checked by visual inspection, to eliminate the 70$\mu$m detection of any SWIRE sources with more than one 24$\mu$m counterpart (3 sources). SWIRE/SDSS sources were then matched with the $MIPS$ 160$\mu$m data, using a search radius of 20$\arcsec$. No sources were found to have a 160$\mu$m detection and more than one 70$\mu$m counterpart. The resulting SWIRE/SDSS spectroscopic catalogue consisted of 1988 sources.

To obtain a low redshift SWIRE/SDSS sample, we choose galaxies with 0.001$\le$z$\le$0.3. Following Bell et al.\ (2003), galactic foreground extinction-corrected 13$\le$$r$$\le$17.5 (Schlegel et al.\ 1998) magnitude limits are applied, since these limits will give a reasonably complete optical sample, as discussed in more detail by e.g., Stoughton et al.\ (2002), Strauss et al.\ (2002) and Blanton et al.\ (2001). 

This produces a final spectroscopic sample of 1114 SWIRE/SDSS galaxies. 

\subsection{Photometry}

Since we are analysing the distribution of SWIRE/SDSS galaxies at low redshift, we require a total measure of optical and infrared flux for extended sources.

For SDSS $ugriz$, we use extinction-corrected Petrosian magnitudes for measuring the total flux of a galaxy, and model magnitudes for measuring colours (see Baldry et al.\ 2004). 

Bright sources which make up our spectroscopic sample have relatively high signal-to-noise ratio measurements of their Petrosian magnitudes, and so are expected to be accurate to better than 0.05 magnitude (Blanton et al. 2003c; Bell et al. 2003; Strauss et al. 2002). These magnitudes recover all of the flux of an exponential galaxy profile and about 80$\%$ of the flux for a de Vaucouleurs profile (Blanton et al.\ 2001). 

For measuring the colours of these sources, we use model magnitudes. These magnitudes are calculated using the best-fit parameters of the light profile (i.e. de Vaucouleurs/exponential) in the $r$-band applied to all other bands. Therefore, the light is measured consistently through the same aperture in all bands, allowing an unbiased comparison of galaxy colours in our sample.

To obtain a measure of the infrared flux of our sample, we use Kron magnitudes for SWIRE 3.6$\mu$m-24$\mu$m bands, since they give the integrated flux through a larger aperture than that used for aperture magnitudes (see Bertin and Arnouts 1996). They are therefore a better measure of total flux, being within $\sim$8$\%$ of the total flux of a galaxy (Kron 1980). As mentioned in $\S$2.1.2, PRF fluxes were used for 70$\mu$m and 160$\mu$m data sets. The difference between PRF fluxes and aperture fluxes for our sample was found to be negligible.

\section{$k$-corrections and luminosities}

To derive rest-frame absolute magnitudes for our sample, we use the SED (Spectral Energy Distribution) template fitting photometric redshift code \renewcommand{\thefootnote}{\fnsymbol{footnote}}$Le$ $Phare$\footnote[2]{\textit{http://www.lam.oamp.fr/arnouts/LE$_{-}$PHARE.html}} (Ilbert et al.\ 2005). Since sources in our sample have spectroscopic redshifts, we only use this code to determine rest-frame magnitudes and for fitting the spectral energy distribution (SED) of sources with galaxy templates.

$Le$ $Phare$ can determine rest-frame magnitudes either by using a measure of apparent magnitude in the same observed band (hereafter $Op0$), or by automatically choosing the apparent magnitude in the observed band closest to the rest-frame band (hereafter $Op1$). For example, at $z$$\ge$0.12 we expect rest-frame $g$-band to be redshifted to observer-frame $r$-band. Similarly, at $z$$\ge$0.08 rest-frame $r$-band will be redshifted to observer-frame $i$-band. Since we are using a direct measurement of the emitted flux in these observed bands instead of the best-fitting SED, this technique limits the dependency on galaxy templates which can be the main source of error and systematics in absolute magnitude measurements.

We test the robustness of our rest-frame magnitude estimates using a simulated 8 square degree catalogue from GALICS (Hatton et al.\ 2003). We extract from the GALICS/MoMaF database 5103 sources with $ugriz$ observer-frame magnitudes, `true' rest-frame magnitudes and redshifts. Applying similar $r$-band magnitude and redshift cuts as for our main spectroscopic sample, gives us a test sample of 678 sources.

Using the 42 synthetic templates of GISSEL 98 (Bruzual and Charlot 1993), we derive $ugriz$ rest-frame magnitudes for our GALICS sample using $Op0$ and $Op1$. We find the dispersion when using $Op0$ to be in the range $\sigma$= 0.027 -- 0.031 mag for $g$-band and $r$-band. However, $Op1$ is found to be in better agreement with the `true' rest-frame magnitudes of GALICS, with a dispersion of 0.025 -- 0.028 mag for $g$-band and $r$-band. 
We therefore adopt $Op1$ for determining rest-frame optical magnitudes for our spectroscopic sample.

We compute the rest-frame luminosities $\nu$$L_{\nu}$ of sources in our sample at 3.6$\mu$m ($L_{3.6}$) and at 24$\mu$m ($L_{24}$). The $Le$ $Phare$ code has been used with a SWIRE library of galaxy and AGN \renewcommand{\thefootnote}{\fnsymbol{footnote}}templates\footnote[3]{\textit{M. Polletta, private communication}} (covering a broad wavelength range from 0.1 to 1000$\mu$m) to fit the optical ($ugriz$) and infrared (3.6$\mu$m, 4.5$\mu$m, 5.8$\mu$m, 8.0$\mu$m and 24$\mu$m) spectral energy distribution of each galaxy, allowing us to determine rest-frame magnitudes at 3.6$\mu$m and 24$\mu$m. We also compute the optical ($L_{OPT}$) and infrared ($L_{IR}$) bolometric luminosities of sources in our sample. Optical luminosities are determined by fitting SWIRE templates to the optical ($ugriz$)+3.6$\mu$m SED of each object, and integrating between optical $u$-band and $z$-band (0.3 -- 0.9$\mu$m). Infrared luminosities are determined by fitting the infrared 5.8 -- 160$\mu$m SED of each object with the full range of Dale $\&$ Helou\ (2002) model template spectra (64 galaxy templates), and integrating between 8 -- 1000$\mu$m.

\section{Spectral Line Analysis}

We use the traditional $\log$([OIII]/$H_{\beta}$) against $\log$([NII]/$H_{\alpha}$) flux ratio BPT  diagram (see e.g. Baldwin, Phillips and Terlevich 1981, Veilleux $\&$ Osterbrock 1987, Kewley et al.\ 2001, and Miller et al.\ 2004) to determine whether galaxies in our sample are star-forming, AGN or systems with quiescent emission. 
Diagnostic diagrams involving [SII]/$H_{\alpha}$ or [OI]/$H_{\alpha}$ flux ratios are considered less effective at determining the level of AGN contamination in star-forming galaxies - see Brinchmann et al.\ (2004). We therefore do not use these ratios in our analysis.

The wavelength separation between the emission lines making up each ratio are small enough that each ratio is relatively insensitive to reddening or flux calibrations. The fluxes of these emission lines are determined using the fitted heights and $\sigma$ of the lines as measured by the SDSS data analysis pipeline.

\begin{figure}
\begin{center}
\includegraphics[height=7cm,width=9cm]{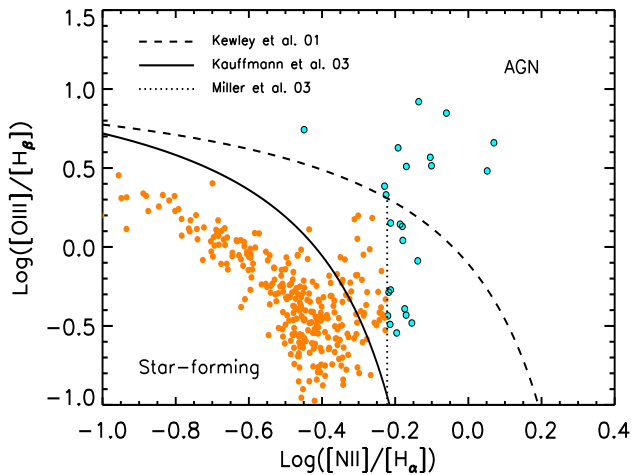}
\caption{BPT (Baldwin, Phillips and Terlevich 1981) diagram in which the emission line flux ratio [OIII]/$H_{\beta}$ versus the ratio [NII]/$H_{\alpha}$ has been plotted for sources with emission line detections above 3$\sigma$ level. Cuts based on the models of Kewley et al.\ (2001) (dashed), Kauffmann et al.\ (2003) (solid) and Miller et al.\ (2004) (dotted) are illustrated. Galaxies are classified as star-forming (orange) or AGN (cyan).}
\end{center}
\end{figure}

Initially, we consider sources with emission lines measurable at above 3$\sigma$ level. Below this limit, a rapidly increasing fraction of galaxies will have a low significance detection in emission or absorption, leading to classification bias (Brinchmann et al.\ 2004). 
Figure 1 shows a diagnostic diagram of 317 out of 1114 galaxies in our spectroscopic sample with emission line detections above 3$\sigma$ level in [OIII], $H_{\beta}$, $H_{\alpha}$, and [NII].

We adopt the emission line classification criterion of Brinchmann et al.\ (2004), using a combination of three cuts to separate star-forming galaxies from AGN in this diagram. Galaxies below the solid line of Kauffmann et al.\ (2003) are expected to be star-forming systems with very low contribution to $H_{\alpha}$ from AGN. We find 257 galaxies lie below this line.
The line of Kewley et al.\ (2001) (dashed) uses a combination of photoionization and stellar population synthesis models to place a conservative lower limit on the number of AGN in our sample. Galaxies with emission line ratios that place them above the line of Kewley et al.\ (2001) cannot be explained by any possible combination of emission line diagnostics that would be characteristic of a star-forming model (Kauffmann et al.\ 2003). 12 galaxies lie above this line, and we therefore classify these as AGN. 

The remaining 48 out of 317 sources lie between the lines of Kauffmann et al.\ (2003) and Kewley et al.\ (2001). These are known as `composite galaxies', since up to 40$\%$ of their $H_{\alpha}$ luminosity may come from an AGN (Brinchmann et al.\ 2004). We therefore use the line of Miller et al.\ (2004) (dotted) in this region to try and classify the remaining composite galaxies as either star-forming or AGN. For these galaxies to have an AGN contamination, they must have [NII] and $H_{\alpha}$ emission lines measurable above the 3$\sigma$ level (regardless of their [OIII] and $H_{\beta}$ emission) and $\log$([NII]/[$H_{\alpha}])$$>$-0.2. 13 composite galaxies meet this criterion and we therefore classify these as AGN. The remaining 35 composite galaxies do not meet any of the requirements for AGN selection, and so we classify these as star-forming systems.

Having classified 317 out 1114 galaxies as star-forming or AGN based on the emission line diagnostic diagram, we attempt to classify the remaining 797 galaxies which do not have emission line detections above 3$\sigma$ accuracy in `all four lines' (i.e., [OIII], $H_{\alpha}$, $H_{\beta}$ and [NII]). 

We again use the emission line criterion for AGN classification of Miller et al.\ (2004) which requires only [NII] and $H_{\alpha}$ lines measurable at above 3$\sigma$ accuracy. We find 162 of these 797 galaxies meet this criterion, and so we classify these as AGN. For the remaining sources to be classified as star-forming systems they must have $H_{\alpha}$ measurable at above 2$\sigma$ accuracy. 214 galaxies meet this criteria. The remaining 421 galaxies could not be classified using emission line diagnostics. This class consists of galaxies with no or very weak emission lines, and are therefore likely to consist of galaxies with quiescent emission.

Having used emission line ratios to classify sources in our sample, we find a total of 187 galaxies to have an AGN component, 506 galaxies to be star-forming systems, and 421 galaxies with quiescent emission. 

\begin{figure*}
\begin{center}
\includegraphics[height=7cm,width=9cm]{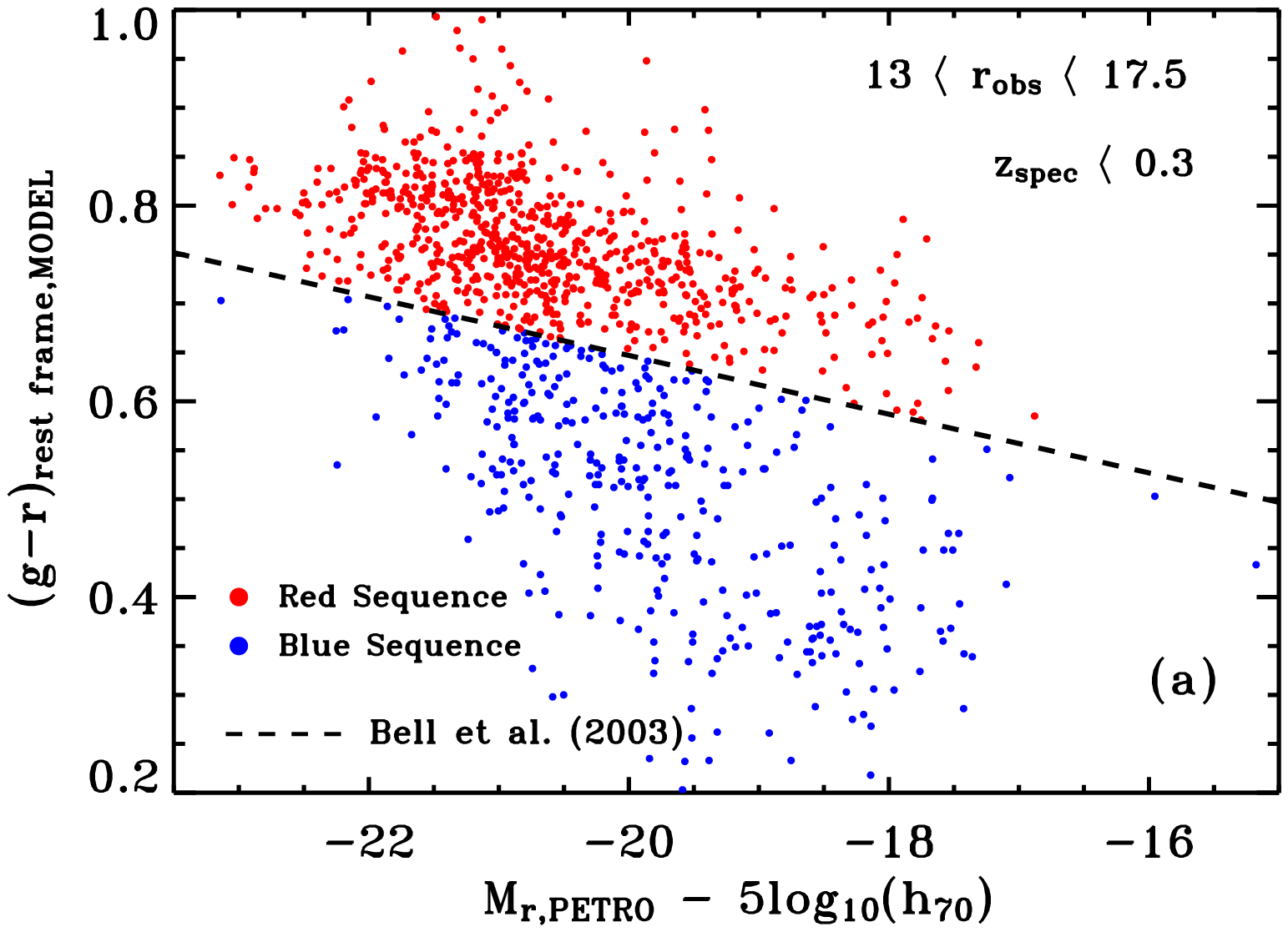}\hspace{-0.5cm}
\includegraphics[height=7cm,width=9cm]{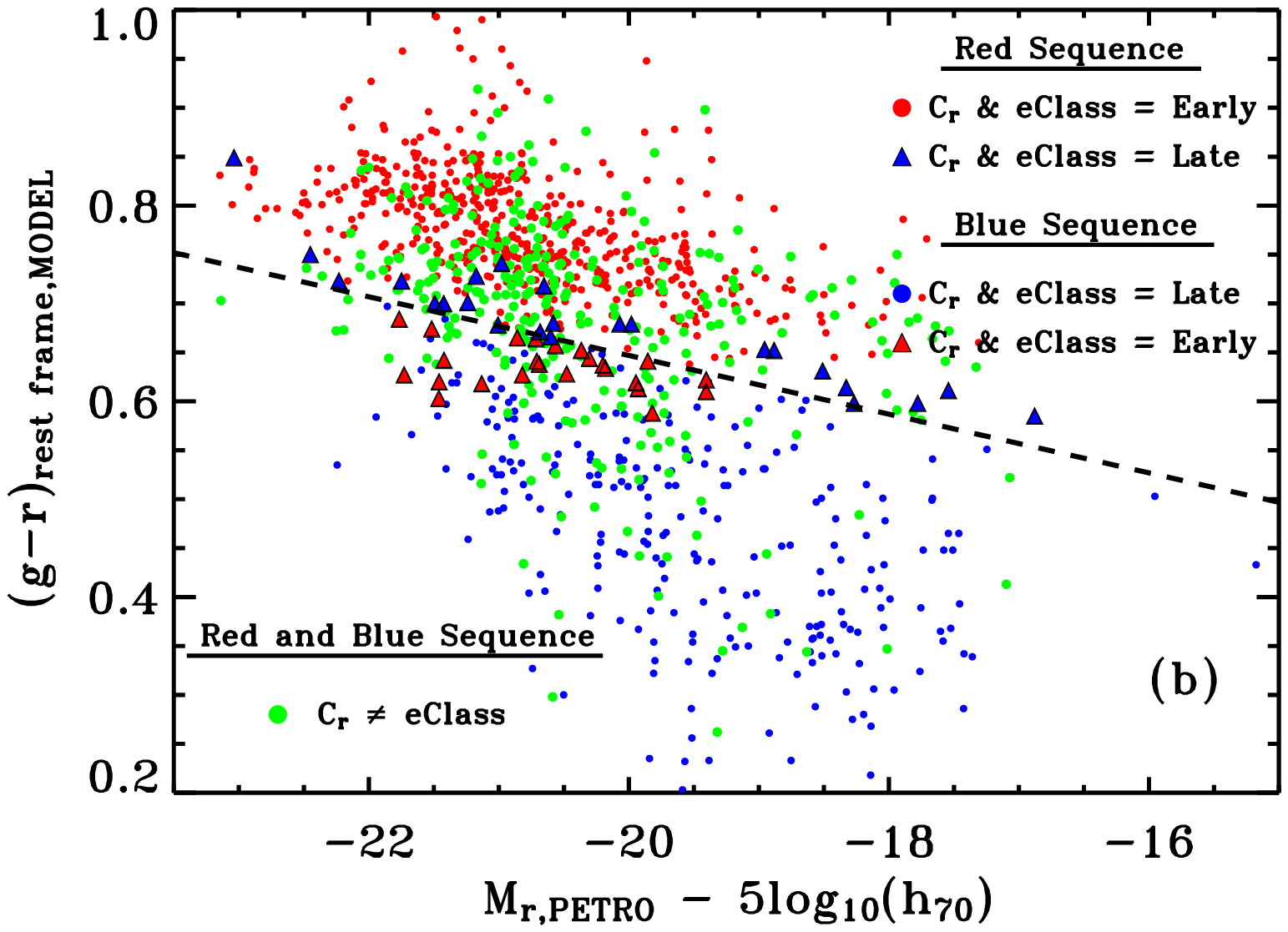}
\includegraphics[height=7cm,width=9cm]{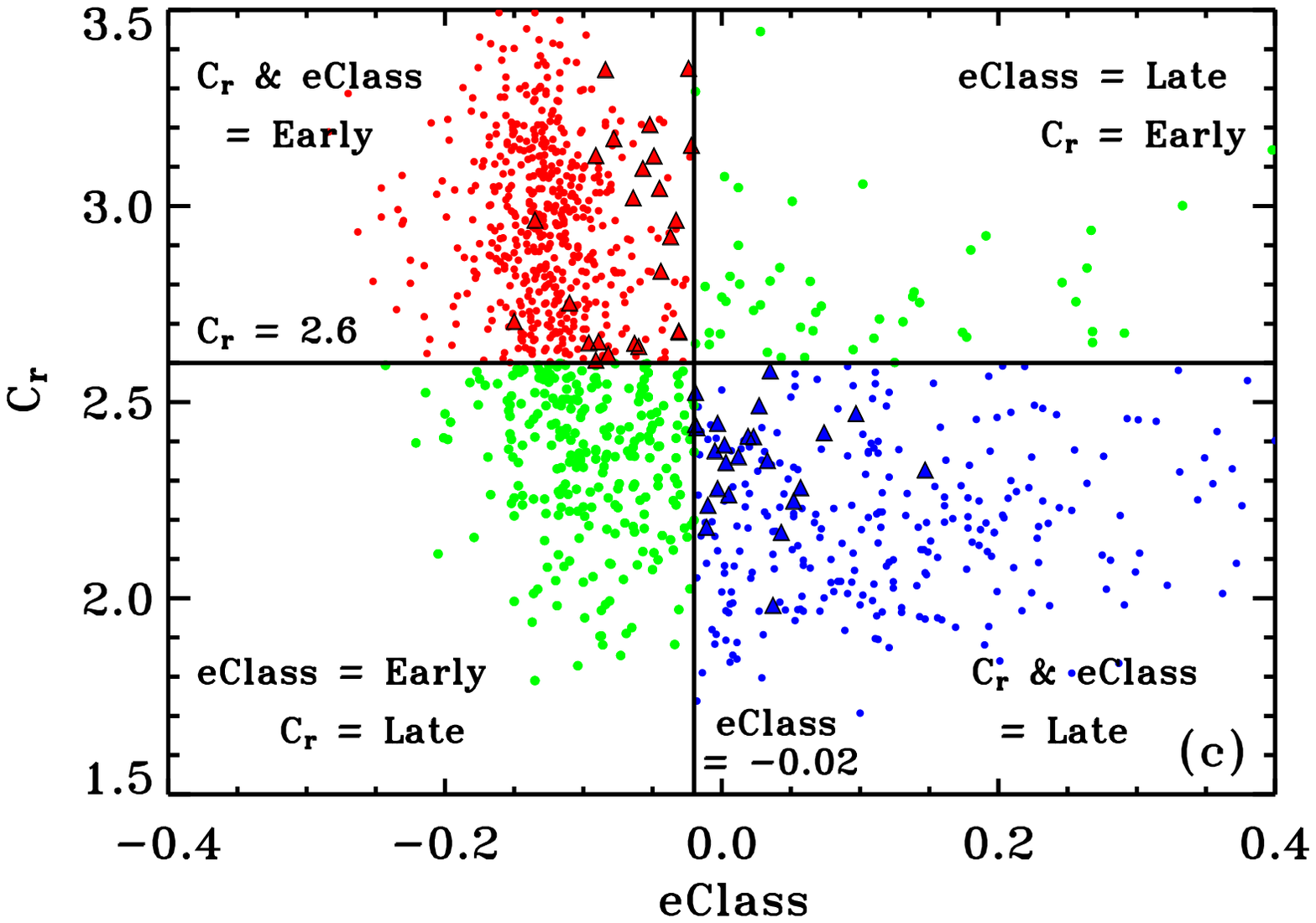}
\hspace{-0.5cm}
\includegraphics[height=7cm,width=9cm]{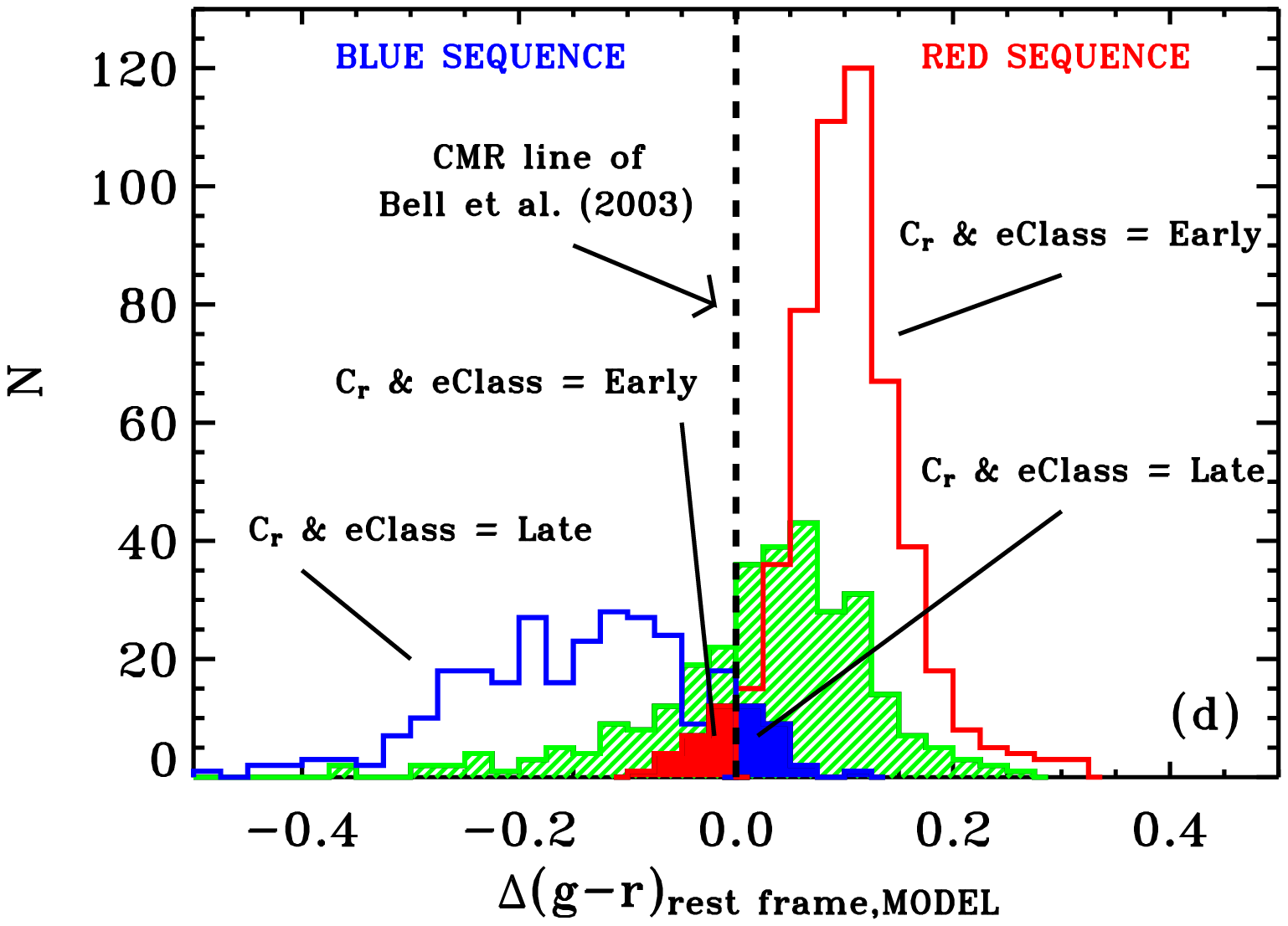}
\caption{The distribution of SWIRE/SDSS galaxy populations; (a) Rest-frame (g-r) colour (magnitudes in AB) of 1114 galaxies against the absolute magnitude in $r$-band, M$_{r,PETRO}$ - 5 log$_{10}$(h$_{70}$). The colour criterion for selection of red sequence and blue sequence galaxies is illustrated by the colour-magnitude line (CMR) of Bell et al.\ (2003) ($dashed$ $line$). (b) Rest-frame (g-r) colour against the absolute magnitude in $r$-band, showing the classification of galaxies according to $C_{r}$ and eClass; red sequence early-type (red filled circle), red sequence late-type (blue filled triangle), blue sequence early-type (red filled triangle), blue sequence late-type (blue filled circle) and disagreement between $C_{r}$ and eClass (green filled circle). (c) Distribution of galaxy populations in terms of $C_{r}$ and eClass. (d) The (g-r) colour of the galaxy populations with respect to the CMR line of Bell et al.\ (2003).}
\end{center}
\end{figure*}

\begin{table*}
\begin{center}
\begin{tabular}{ c c c c c c c c c }
\hline\hline
SEQUENCE & No. & \multicolumn{2}{c}{$C_{r}$} & \multicolumn{2}{c}{eclass} & \multicolumn{2}{c}{$C_{r}$} {=}  {eClass} & {$C_{r}$} {$\neq$} eClass\\
\\
 & & $early$ & $late$ & $early$ & $late$ & $early$ & $late$ &\\
 \cline{1-9}
 \\
RED & 741 & 70$\%$ & 30$\%$ & 95$\%$ & 5$\%$ & 69$\%$ & 3$\%$ & 28$\%$\\
BLUE & 373 & 16$\%$ & 84$\%$ & 22$\%$ & 78$\%$ & 6$\%$ & 68$\%$ & 26$\%$\\
\hline
\end{tabular}\\
\end{center}
\caption{SDSS morphological (concentration index $C_{r}$) and spectral classification (eClass) for sources in the red and blue sequence.}
\end{table*}

\section{Optical Colour-Magnitude Relation}

Figure 2a illustrates rest-frame (g-r) colour as a function of $r$-band absolute magnitude for our spectroscopic sample. Red sequence early-type galaxies and blue sequence late-type galaxies are initially defined by adopting the colour-magnitude relation (CMR) line of Bell et al.\  (2003). We find two-thirds of our sample contain galaxies with red optical colours and the remaining third have blue optical colours (Table 1). 

We use SDSS morphological ($C_{r}$) and spectral (eClass) classifiers to determine the fraction of early and late-type galaxies in each sequence (Figures 2b and 2c). The distribution of our galaxy sample is found to show a continuous trend in classifiers $C_{r}$ and eClass (Figure 2c). 

Approximately 70$\%$ of red sequence galaxies are defined as early-type according to both $C_{r}$ and eClass (red filled circle). Similarly, 70$\%$ of blue sequence galaxies are defined as late-type (blue filled circle). As illustrated by Figure 2b, only 3$\%$ of galaxies in the red sequence are late-type (blue triangle) and 6$\%$ of galaxies in the blue sequence are defined as early-type (red triangle) according to both classifiers. These galaxies make up less than 4$\%$ of the entire spectroscopic sample, and also lie near the boundaries of the classifiers $C_{r}$ and eClass (Figure 2c). As shown by Figure 2d, galaxies in the blue sequence classified as early-type are likely to correspond to the tail of the early-type distribution in the red sequence. Likewise, the late-type galaxies in the red sequence correspond to the tail of the late-type distribution in the blue sequence. In terms of their (g-r) colour, these sources lie in a `transition region' between the two sequences (Figures 2b and 2d), either side of the CMR line of Bell et al.\ (2003). 

Therefore, the CMR line is found to correspond well to the boundaries of $C_{r}$ and eClass, and can be used as a reliable separator of red sequence galaxies dominated by early-types and blue sequence galaxies dominated by late-types.

In Figure 2c, galaxies classified as early-type (red) and late-type (blue) are found to be separated by a population where the two classifiers $C_{r}$ and eClass disagree (green circle). These sources make up $\sim$27$\%$ of the SWIRE/SDSS sample, the majority of which are classified as early-type according to eClass and late-type according to $C_{r}$. These sources are distributed throughout both sequences, with the majority (19$\%$ of the SWIRE/SDSS sample) located in the red sequence. As shown in Figure 2d, they have slightly bluer (g-r) colour than early-type galaxies in the red sequence and a broader distribution (with $\sigma_{green}$ $\sim$ 0.1), with a skewed tail which passes through most of the blue sequence. The percentage of these sources found in each sequence is given in Table 1.

To gain another viewpoint of the galaxy populations in our sample, we look at their spectral classifications (Figure 3) according to the optical spectral line analysis discussed in $\S$4. Star-forming systems are found to dominate the blue sequence, while passive galaxy populations dominate the red sequence. However, a significant number of AGN and star-forming galaxies are also found in the red sequence. This may explain why a significant population of galaxies in the red sequence have been classified as eClass early-type and $C_{r}$ late-type, as illustrated by Figure 2.

By investigating the infrared colours of optically selected red and blue sequence galaxies in our sample, we can therefore understand the nature of these active systems in the red sequence, aswell as the origin of their red optical colours (i.e. old stellar populations or extinction). Such an infrared analysis will also help determine whether the bi-modal distribution seen in the optical is too simplistic in its representation of early and late-type galaxies in the local universe.

\section{The Infrared Colour Distribution}

\begin{figure}
\begin{center}
\includegraphics[height=7cm,width=9cm]{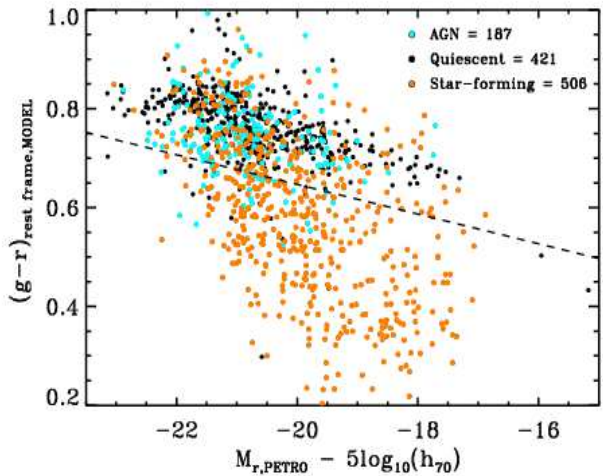}
\caption{Classification of sources according to the spectral line analysis in $\S$4; AGN (cyan), Star-forming (orange) and galaxies with quiescent emission (black).}
\end{center}
\end{figure}

Starlight at 3.6$\mu$m can be used to trace the stellar mass distribution of galaxies almost free of dust obscuration effects (i.e. Oliver et al.\ 2004; Pahre et al.\ 2004; Franceschini et al.\ 2006), since any standard extinction law predicts only a small percentage of extinction compared to optical wavelengths. 

On the other hand, the study of M51a using $Spitzer$ and $HST$ data has shown directly that 24$\mu$m luminosity can be used as a reliable indicator of obscured star-formation (i.e. Calzetti et al.\ 2005). 24$\mu$m observations of individual HII regions were calibrated against Pa$\alpha$ (1.876$\mu$m) observations, which measure the ionized radiation from stars forming in the arms of M51a. A tight linear correlation was found between Pa$\alpha$ and 24$\mu$m, proving that the 24$\mu$m dust continuum is a locally accurate tracer of star-formation. In addition, the goodness of the mid-infrared emission as a star-formation tracer at high redshift has been found recently by Marcillac et al.\ (2006), testing the mid-infrared emission against the radio emission of galaxies (see also Elbaz et al.\ 2002; Gruppioni et al.\ 2003; Appleton et al.\ 2004).

We therefore use and interpret infrared 3.6$\mu$m luminosity as an indicator of stellar mass and 24$\mu$m as an indicator of obscured star-formation.

\begin{figure*}
\begin{center}
\includegraphics[height=7cm,width=9cm]{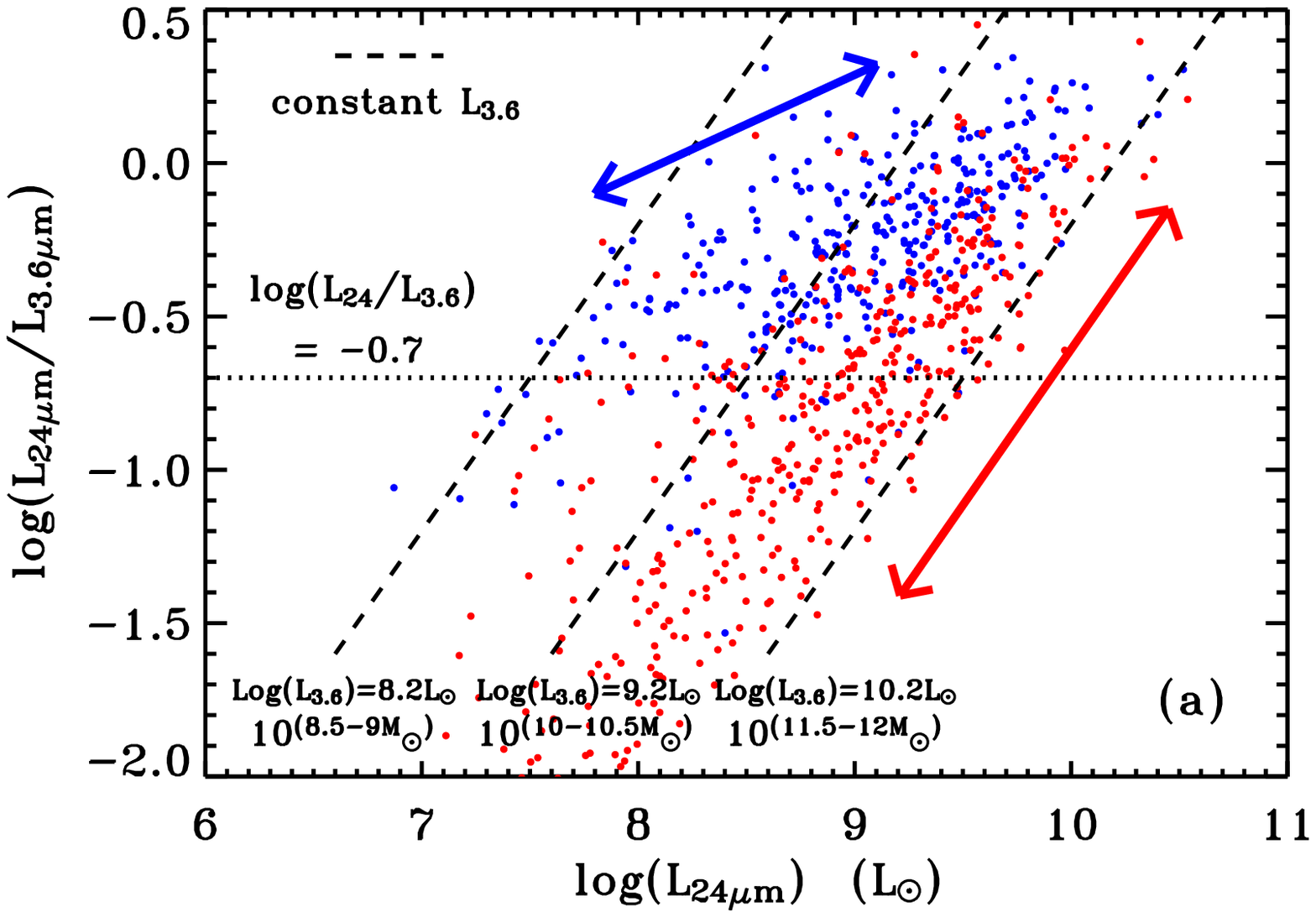}\hspace{-0.5cm}
\includegraphics[height=7cm,width=9cm]{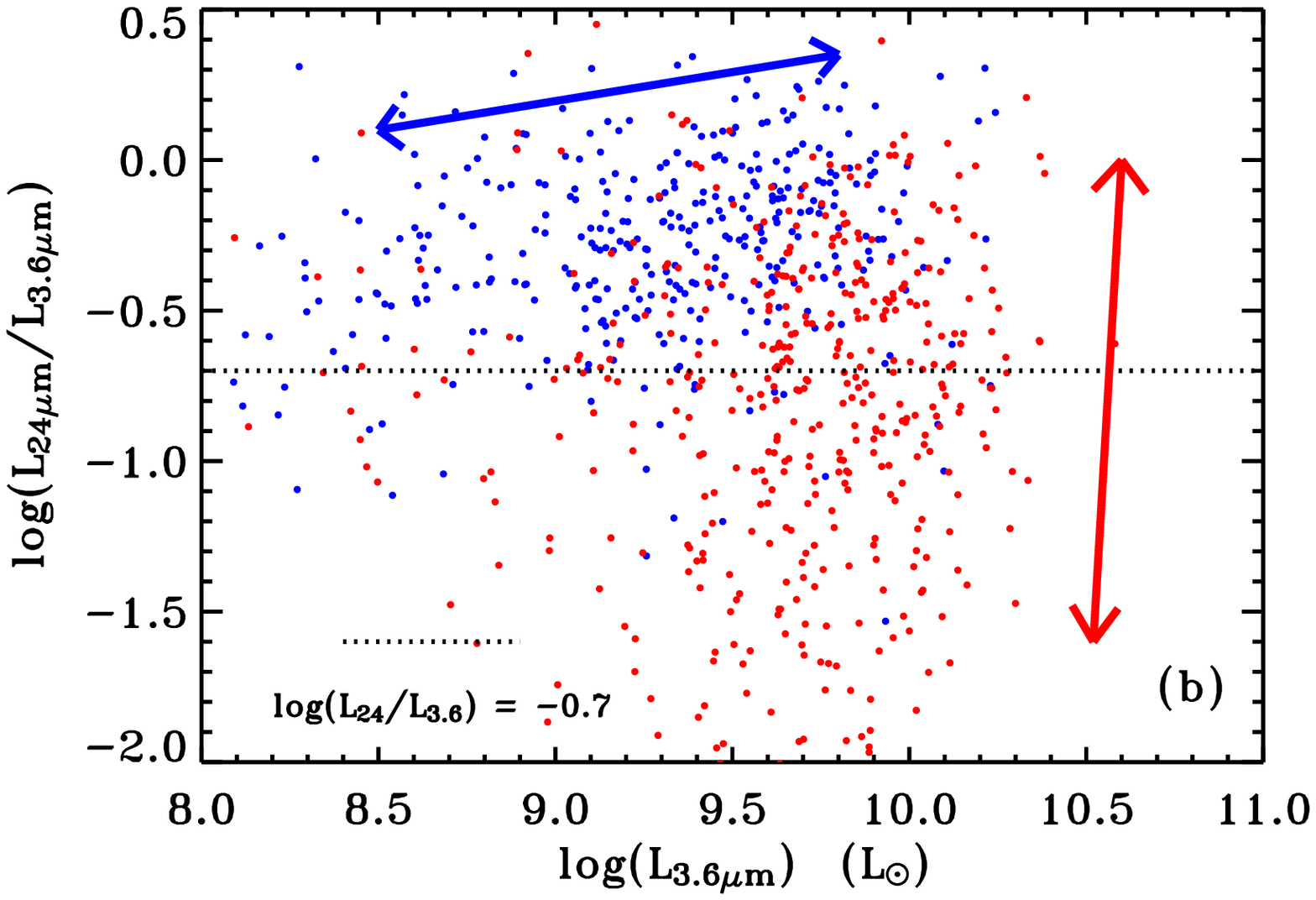}
\caption{Ratio of 24$\mu$m to 3.6$\mu$m luminosity as a function of (a) 24$\mu$m luminosity and (b) 3.6$\mu$m luminosity. Luminosities are in $\nu$$L_{\nu}$. Colours are coded according to Figure 2a, where galaxies above the CMR line of Bell et al.\ (2003) are in red and galaxies below the line are in blue (see Figure 2a). A cut has been applied at $\log$($L_{24}$/$L_{3.6}$) = -0.7 (dotted line), above which galaxies are dominated by active systems. Dashed lines in Figure 4a signify regions of constant 3.6$\mu$m luminosity. The arrows show the trend of each sequence.}
\end{center}
\end{figure*}
\begin{figure*}
\begin{center}
\includegraphics[height=7cm,width=9cm]{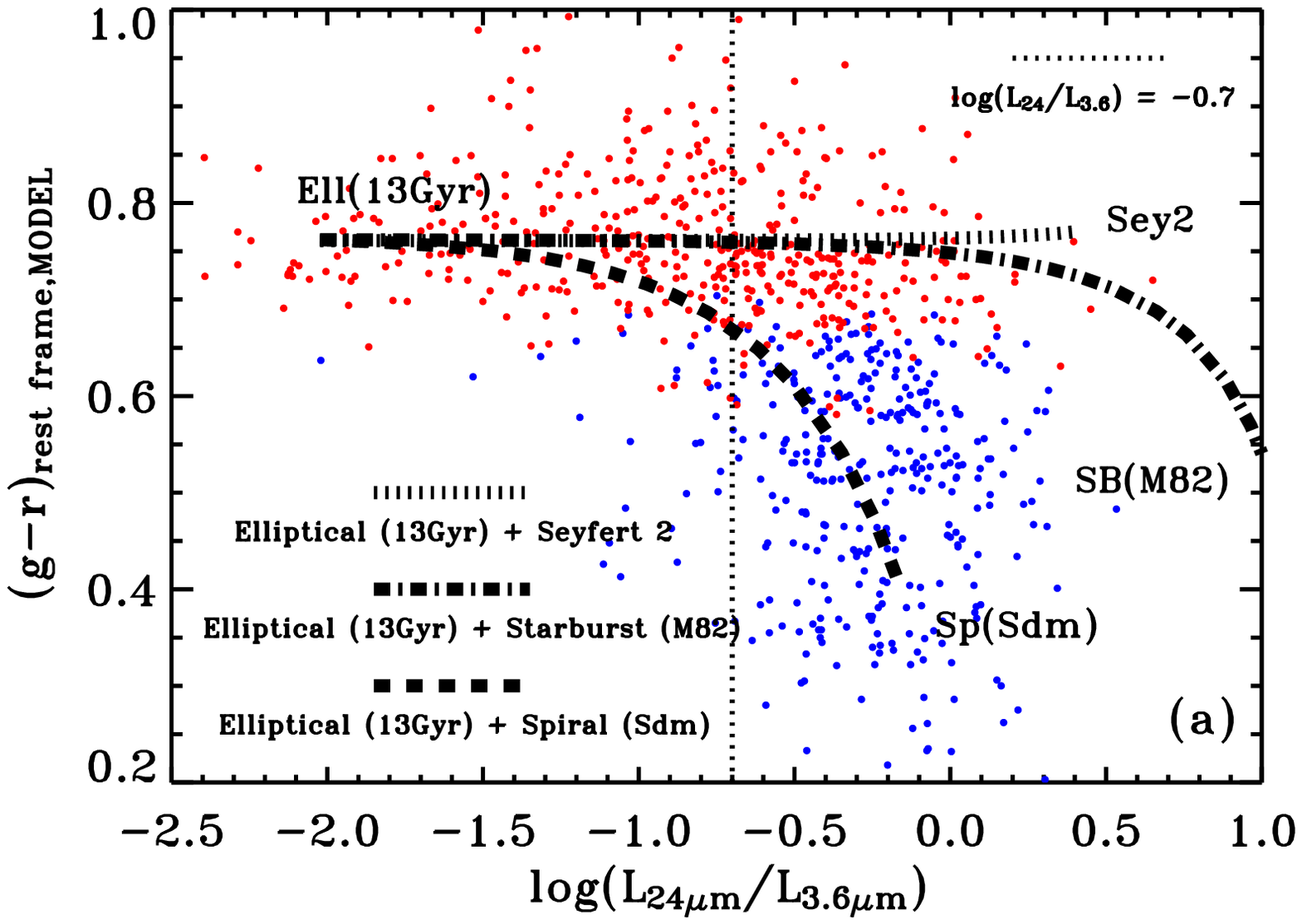}\hspace{-0.5cm}
\includegraphics[height=7cm,width=9cm]{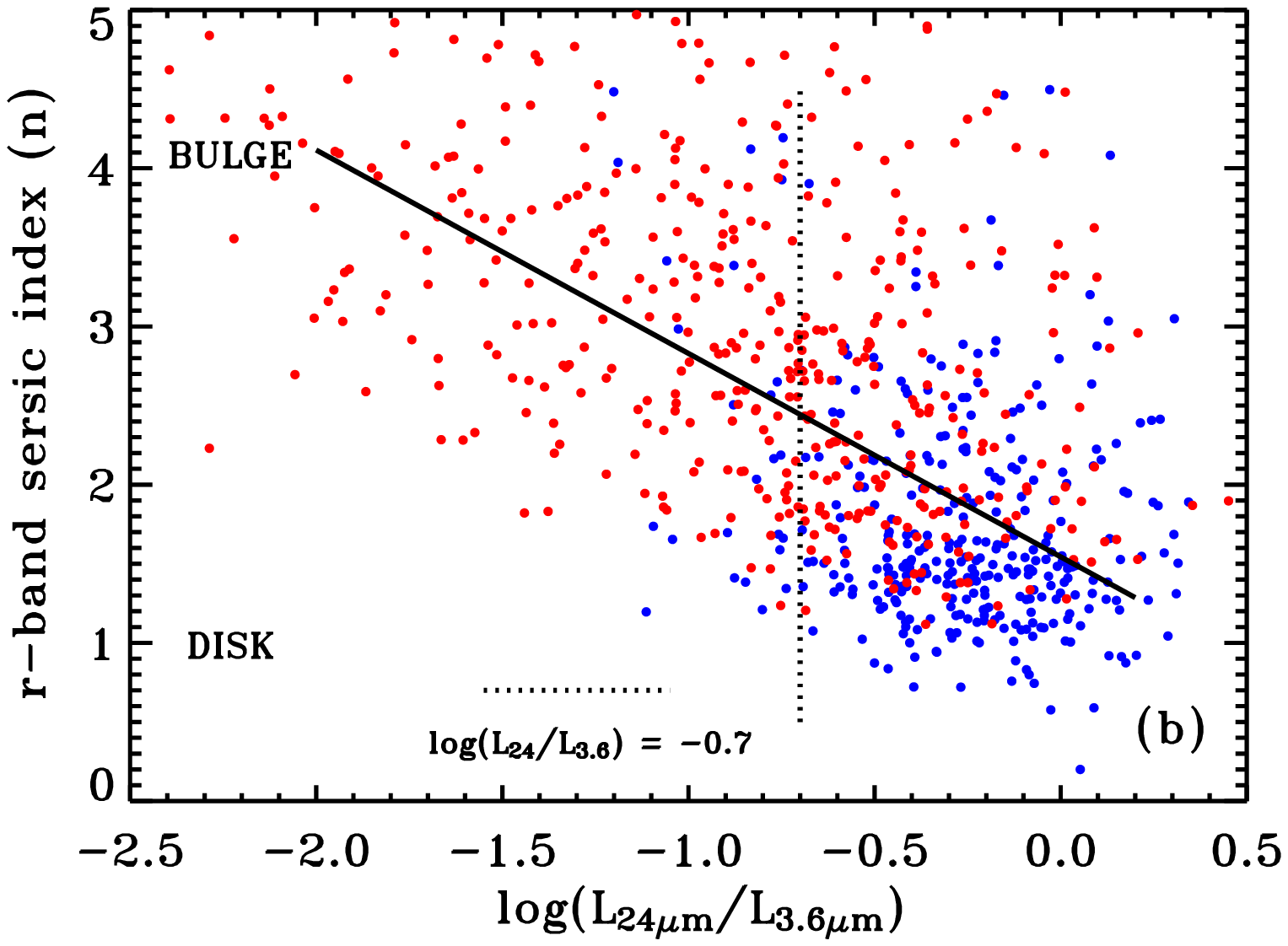}
\caption{(a) Rest-frame (g-r) colour against the ratio of 24$\mu$m to 3.6$\mu$m luminosity. Illustrated are red and blue sequence galaxies with galaxy template tracks overplotted; Dotted-dashed track = Elliptical (13Gyrs) + Starburst (M82) template, dashed track = Elliptical (13Gyr) + Spiral (Sdm) template, dotted track = Elliptical (13Gyr) + Seyfert 2 template. The tracks show that the disk-like contribution from the late-type templates (Spiral, Starburst, Seyfert 2) increase with $\log$($L_{24}$/$L_{3.6}$) ratio, while the bulge-like contribution from the early-type template (Elliptical) decreases. (b) SDSS $r$-band S$\acute{{e}}$rsic index against the ratio of 24$\mu$m to 3.6$\mu$m luminosity. A linear fit to the distribution of the data is illustrated (solid line).}
\end{center}
\end{figure*}

In Figure 4 the ratio $\log$($L_{24}$/$L_{3.6}$) as a function of $\log$($L_{24}$) (Figure 4a) and $\log$($L_{3.6}$) (Figure 4b) are shown for sources in our sample with 24$\mu$m detections (787 sources out of 1114, $\sim$70$\%$ of our sample). $L_{24}$ and $L_{3.6}$ are the luminosities ($\nu$$L_{\nu}$ in L$_{\odot}$) at 24$\mu$m and 3.6$\mu$m. Regions of constant stellar mass are illustrated in Figure 4a by straight dashed lines of 3.6$\mu$m luminosity. The stellar mass approximations shown in Figure 4a are based on the SDSS stellar mass catalogue of Kauffmann et al.\ (2003).

\begin{table}
\caption{Fraction of red sequence and blue sequence galaxies with $\log$($L_{24}$/$L_{3.6}$) $>$ -0.7 and $\log$($L_{24}$/$L_{3.6}$) $<$ -0.7.}
\begin{center}
\begin{tabular}{c c c c c}
\hline\hline
SEQ. & No. & 24$\mu$m det. & $\log$($L_{24}$/$L_{3.6}$) & $\log$($L_{24}$/$L_{3.6}$)  \\
& & & $>$ -0.7 & $<$ -0.7  \\
 \cline{1-5}
RED & 741 & 440 & 195 & 245 \\
BLUE & 373 & 347 & 317 & 30 \\
\hline
\end{tabular}\\
\end{center}
\end{table}

For both red and blue sequence galaxies, the ratio $\log$($L_{24}$/$L_{3.6}$) increases as a function of $\log$($L_{24}$). A similar relation between mid-infrared (15$\mu$m) and optical luminosities was found to be valid for ISOCAM sources (see Pozzi et al.\ 2004; La Franca et al.\ 2004 and Rowan-Robinson et al.\ 2005). In Figure 4b we see that the $\log$($L_{24}$/$L_{3.6}$) ratio of blue sequence galaxies is found to increase with 24$\mu$m luminosity, but $\log$($L_{24}$/$L_{3.6}$) shows little variation with 3.6$\mu$m luminosity. Since $L_{3.6}$ is indicative of stellar mass and $L_{24}$ of star-formation rate, the ratio of $\log$($L_{24}$/$L_{3.6}$) can be interpreted as an indication of the level of star-formation activity per unit stellar mass or ``specific star-formation" (see also Pozzi et al.\ 2004 and Rowan-Robinson et al.\ 2005). We can then interpret Figures 4a and 4b as being determined primarily by two almost independent factors; the stellar mass and the specific star-formation.  

For red sequence galaxies the specific star-formation is independent of stellar mass. This is clearly seen in Figure 4b where they cluster around $\log(L_{3.6})\sim9.8$$L_{\odot}$ but with a wide range of $\log$($L_{24}/L_{3.6}$), and in Figure 4a where the distribution runs parallel to lines of constant stellar mass.  The blue sequence galaxies on the other hand appear to have a comparatively constant specific star-formation activity, though there is a slight tendency for more massive blue galaxies to have higher specific star-formation.  This weak trend is in the opposite sense from trends seen in the optical and indicates a more complex relation between the star formation indicators, which we will explore later.

Note that the star-formation in red sequence galaxies with very low $\log$($L_{24}$/$L_{3.6}$) ratio is likely to correspond to quiescent rather than active star-formation, due to the interaction between dust and the old stellar populations that exist in these galaxies (Wang \& Heckman 1996). 

Looking more at luminosities than colours, both red and blue sequence galaxies are found to have similar ranges of 24$\mu$m luminosity ($L_{24} = 10^{7}-10^{10} L_{\odot}$). Red sequence galaxies are found to have higher 3.6$\mu$m luminosities than galaxies in the blue sequence, which would suggest that galaxies in the red sequence generally have higher stellar mass ($10^{10}-10^{12}M_{\odot}$) than galaxies in the blue sequence ($10^{8.5}-10^{11}M_{\odot}$).

To separate galaxies with young and evolved star-formation, we adopt a cut at $\log$($L_{24}$/$L_{3.6}$) = -0.7, below which the number of blue sequence galaxies is negligible (see Figure 4a and 4b).
Considering sources with 24$\mu$m detections, 92$\%$ of galaxies in the blue sequence are above this cut (Table 2), consistent with their nature as actively star-forming systems. However, 44$\%$ of red sequence galaxies (195 sources) also have $\log$($L_{24}$/$L_{3.6}$)$>$-0.7, unexpected of optically red quiescent emitting galaxies. 

To make a direct comparison of the optical and infrared colours of red and blue sequence galaxies in our sample, we illustrates the ratio $\log$($L_{24}$/$L_{3.6}$) as a function of rest-frame (g-r) colour (Figure 5a). 
The (g-r) colour of red sequence galaxies is found to remain constant with increasing $\log$($L_{24}$/$L_{3.6}$) ratio. Red sequence galaxies also have a small spread in (g-r) colour, with $\sigma_{(g-r)}$$\sim$0.07. Above our adopted cut of $\log$($L_{24}$/$L_{3.6}$)=-0.7, blue sequence galaxies dominate, and have a large range of (g-r) colour than red sequence galaxies, with $\sigma_{(g-r)}$$\sim$0.12. 

To gain a better understanding of the distribution of red and blue sequence galaxies in optical and infrared colours, galaxy template tracks are overplotted in Figure 5a. The tracks used are from the SWIRE template library. Illustrated are a combination of (i) a 13Gyr Elliptical template and a Starburst (M82) template (dotted-dashed), (ii) a 13Gyr Elliptical template and a Spiral (Sdm) template (dashed) and (iii) a 13Gyr Elliptical template and a Seyfert 2 template (dotted). 

Galaxies with low $\log$($L_{24}$/$L_{3.6}$) ratio and red optical colours are dominated by an Elliptical-like contribution. These galaxies are bulge dominated systems with quiescent star-formation activity. 

As the ratio $\log$($L_{24}$/$L_{3.6}$) increases, the contribution from the late-type templates increases at varying rates. Galaxies with blue optical colours are completely disk dominated systems where the optical and infrared emission comes from a spiral galaxy. 
Galaxies with red optical colours and $\log$($L_{24}$/$L_{3.6}$)$>$-0.7 are found to have up to an 80$\%$ AGN or 10$\%$ starburst contribution. Galaxies with an AGN owe their red optical colours to their bulge component and their high $\log$($L_{24}$/$L_{3.6}$) ratio to dust heated by the AGN. Extinction could account for the red optical colours of galaxies with a starburst component. However, we find that these systems can have up to a 90$\%$ Elliptical-like contribution, where their red optical colours could be due to a significant bulge component (i.e. old rather than extincted stars). Since these systems can have up to a 10$\%$ starburst contribution, the high $\log$($L_{24}$/$L_{3.6}$) ratio of these galaxies would correspond to intense bursts of star-formation activity. Analysing the optical and infrared colours of galaxies in our sample, we therefore observe a possible trend in bulge-to-disk ratio as a function of $\log$($L_{24}$/$L_{3.6}$) colour. 

To test whether there is a relationship between the morphology and the infrared emission of galaxies in our sample, we analyse their bulge-to-disk ratio (S$\acute{{e}}$rsic index - Sersic 1968) as a function of infrared $\log$($L_{24}$/$L_{3.6}$) colour (Figure 5b). We use the SDSS $r$-band S$\acute{{e}}$rsic indices (see $\S$2.3) as measured by Blanton et al.\ (2003c) for SDSS galaxies. 

A correlation is found between S$\acute{{e}}$rsic index and the infrared colours of galaxies. Rojas et al.\ (2004) find that galaxies with $n$ $<$1.8 have surface brightness profiles that resemble spiral galaxies, while galaxies with $n$ $>$ 1.8 are more like ellipticals. We find that red sequence galaxies with low $\log$($L_{24}$/$L_{3.6}$) ratio have a high S$\acute{{e}}$rsic index ($n$$>$2) and are bulge dominated early-type systems. The majority of galaxies with $\log$($L_{24}$/$L_{3.6}$)$>$-0.7 are found to have a low S$\acute{{e}}$rsic index ($n$$<$1.5), and are blue sequence disk-dominated systems. However, red sequence galaxies with $\log$($L_{24}$/$L_{3.6}$)$>$-0.7 have a S$\acute{{e}}$rsic index $>$1.5, indicating that they are bulge-dominated. For galaxies with an AGN contribution based on the spectral analysis, the bulge component would account for the high S$\acute{{e}}$rsic index. For galaxies with only a starburst component based on the spectral analysis, the red optical colours could be accounted for by a bulge component, and the high mid-infrared emission characterised by bursts of star-formation from a disk component.

\section{Red Sequence galaxies with high infrared emission}

Galaxies exhibiting red optical colours and high infrared luminosities can be considered either AGN or massive dust-enshrouded starburst galaxies (Bell et al.\ 2005). Based on the spectral classifications of our sample (see $\S$4), 78 of the 195 red sequence galaxies with $\log$($L_{24}$/$L_{3.6}$)$>$-0.7 contain an AGN component, and 117 galaxies are star-forming systems. 

In this section we further investigate this sub-sample of galaxies, to determine whether dust-obscuration is responsible for the red optical colours of the star-forming galaxies, and whether the integrated optical and infrared luminosities of these systems differ from optically red quiescent galaxies and star-forming galaxies with blue optical colours.

\subsection{Optical extinction of star-forming galaxies}

We have already corrected our optical magnitudes for galactic foreground obscuration when determining rest-frame optical colours (see Figure 2a). However, obscuration by dust intrinsic to star-forming galaxies can affect the optical colours of a galaxy. 

\begin{figure}
\begin{center}
\includegraphics[height=7cm,width=9cm]{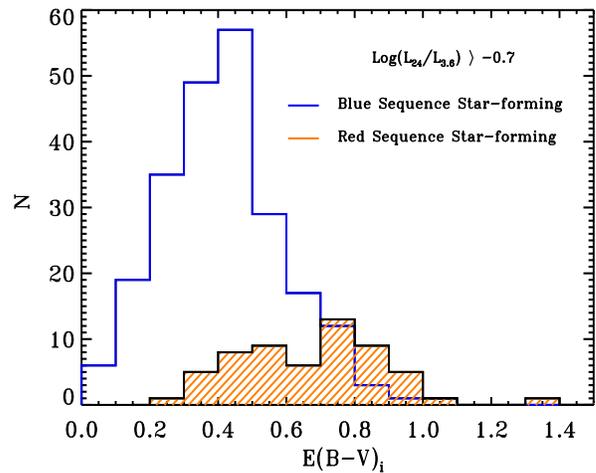}
\caption{Colour excess $E(B-V)_{i}$ of red sequence (orange) and blue sequence (blue) star-forming galaxies with $\log$($L_{24}$/$L_{3.6}$)$>$-0.7}
\end{center}
\end{figure}

\begin{table}
\caption{Median colour excess $E(B-V)_{i}$ of star-forming galaxies with $\log$($L_{24}$/$L_{3.6}$)$>$-0.7.}
\begin{center}
\begin{tabular}{c c c}
\hline\hline
SEQ. & Median $E(B-V)_{i}$ & St.Dev.  \\
& & \\
 \cline{1-3}
RED & 0.70 & 0.27\\
BLUE & 0.41 & 0.19\\
\hline
\end{tabular}\\
\end{center}
\end{table}

At optical wavelengths, the Balmer decrement (consisting of Balmer line ratios $H_{\alpha}$/$H_{\beta}$) can provide a measure of extinction for optical spectra due to intrinsic reddening (down to an optical depth $\tau$$\sim$1), in galaxy regions where bursts of star-formation are located (Ward et al.\ 1987, Calzetti et al.\ 2004). 

To determine the Balmer Decrement, we first apply corrections for stellar absorption to our $H_{\alpha}$ and $H_{\beta}$ flux measurements. Balmer emission lines sit on top of stellar absorption due to the presence of young and intermediate age stars in the line-emitting galaxy. Therefore, Balmer emission lines uncorrected for this absorption can cause an overestimation in the obscuration measurements derived from the Balmer Decrement (Hopkins et al.\ 2003).

To obtain accurate measures of extinction, we consider galaxies with $H_{\alpha}$ and $H_{\beta}$ emission greater than 3$\sigma$ accuracy. For sources with $H_{\alpha}$ $>$ 3$\sigma$ accuracy, but with $H_{\beta}$ $<$ 3$\sigma$, we set the $H_{\beta}$ flux to its 3$\sigma$ error value. This will allow us to obtain a lower limit measure of the extinction of these sources (see e.g., Babbedge et al.\ 2004).

The intrinsic colour excess $E(B-V)_{i}$ for a uniform screen of dust is related to the Balmer line ratio $H_{\alpha}$/$H_{\beta}$, according to (e.g. Hummer \& Storey 1987, Calzetti et al.\ 1994):

\begin{equation}
E(B-V)_{i} \hspace{0.05cm} {\approx} \hspace{0.05cm} 0.935 \hspace{0.05cm} ln \left(\frac{H_{\alpha}/H_{\beta}}{2.88}\right)
\end{equation}

We compare the intrinsic colour excess $E(B-V)_{i}$ of all star-forming galaxies in our sample with $\log$($L_{24}$/$L_{3.6}$)$>$-0.7; 117 red sequence galaxies and 281 blue sequence galaxies. 
Figure 6 shows $E(B-V)_{i}$ for our two populations of star-forming galaxies, with median values given in Table 3.

After applying an extinction correction to the star-forming galaxies in the red sequence, we find $\sim$51$\%$ of the sample (60 sources) now have optical colours consistent with blue sequence galaxies. These sources are found to have an extinction of $E(B-V)_{i}$$\ge$0.3, similar to previous mid-infrared selected sources (i.e. Poggianti and Wu 2000; Pozzi et al.\ 2003). An $E(B-V)_{i}$$<$0.3 is not found to be enough to change the optical colours of these sources. 
The remaining 49$\%$ of the sample (57 sources) either have $E(B-V)_{i}$$<$0.3 or a value of $E(B-V)_{i}$ could not be determined because $H_{\alpha}$ did not meet the 3$\sigma$ emission criteria. This would indicate that these remaining systems are either so extinct that their Balmer lines have been surpressed, or that they are early-type systems characterised by weak Balmer emission.

\begin{figure}
\begin{center}
\includegraphics[height=7cm,width=9cm]{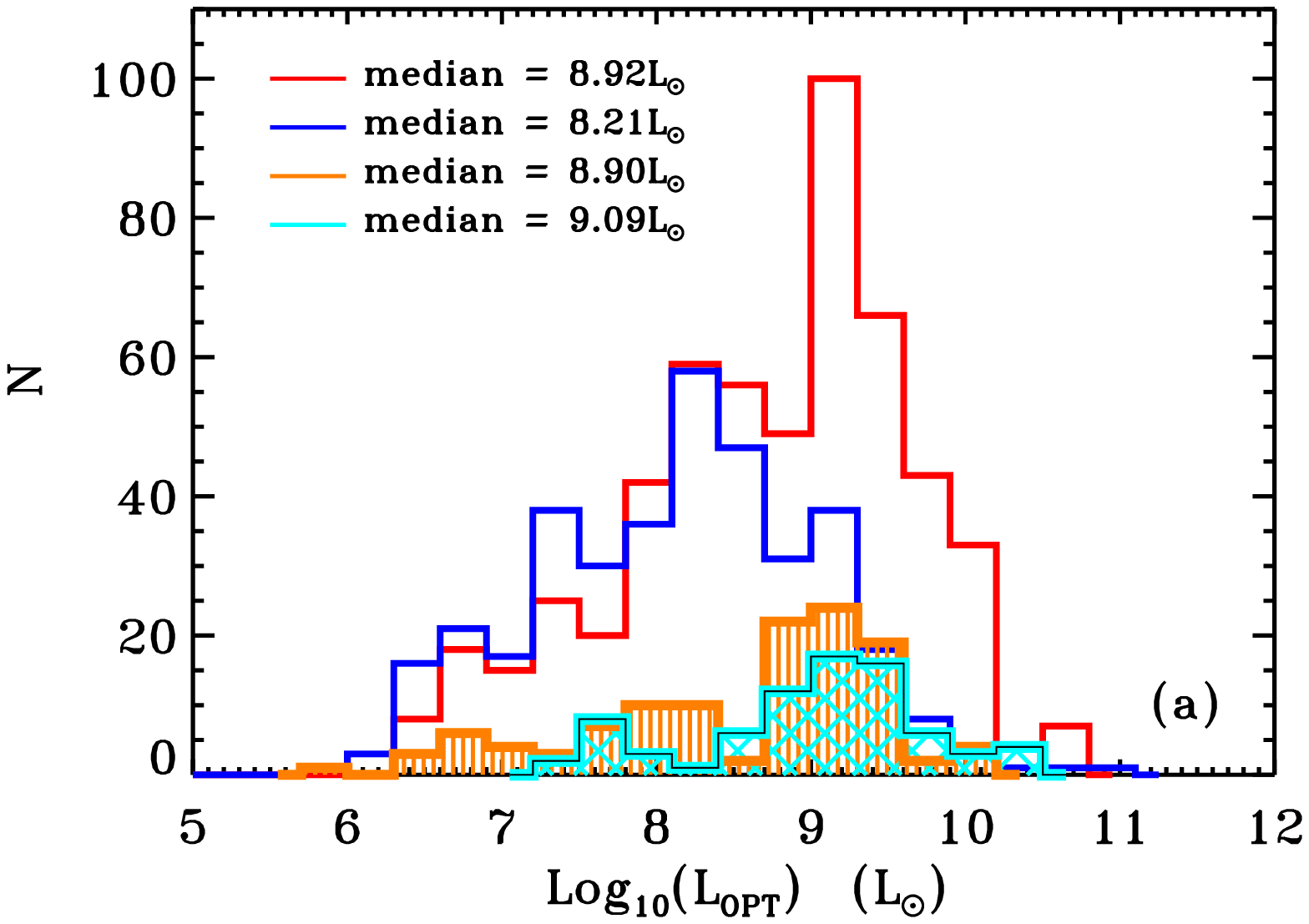}
\includegraphics[height=7cm,width=9cm]{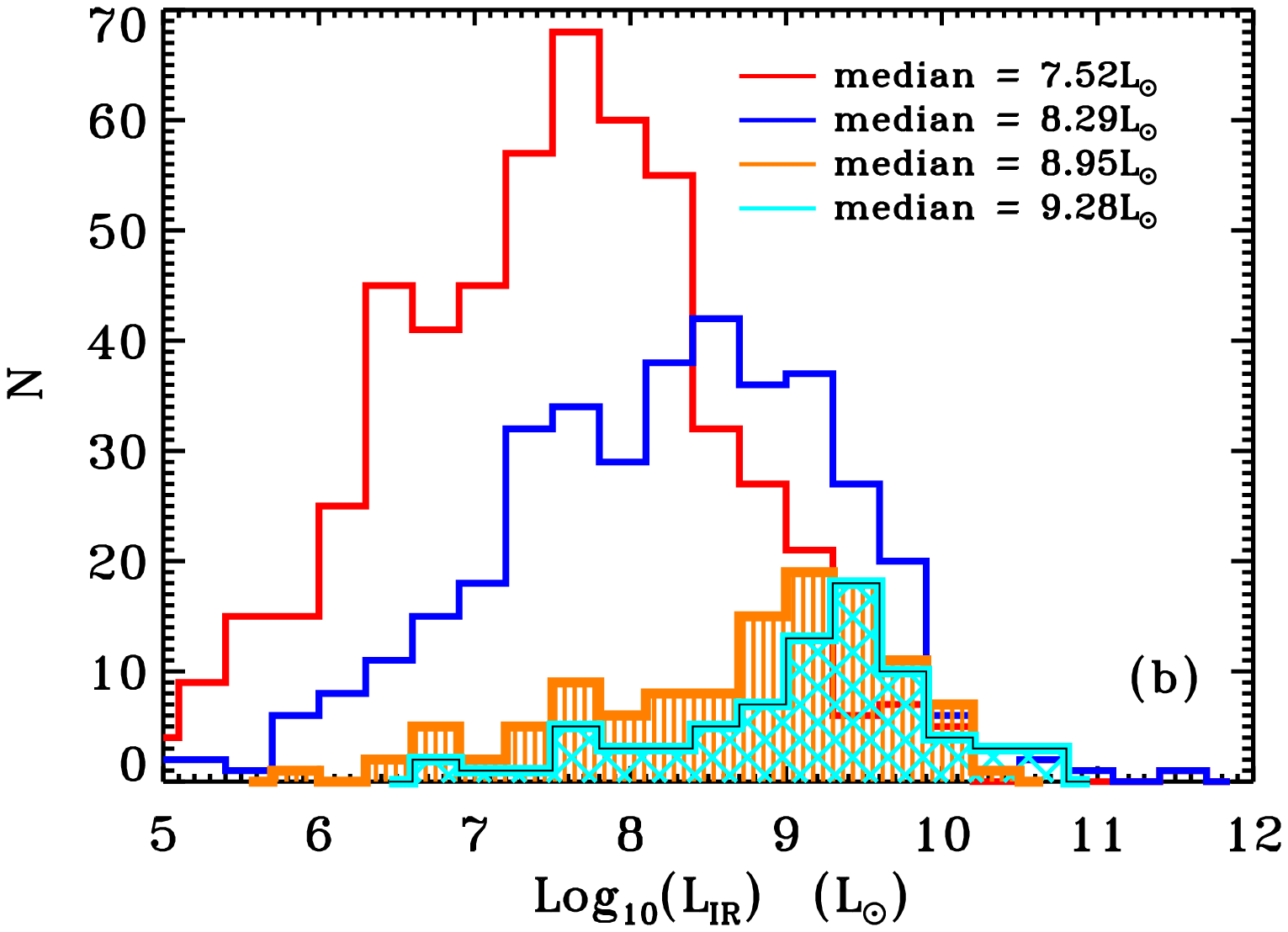}
\caption{(a) Optical (L$_{\rm OPT}$) and (b) infrared (L$_{\rm IR}$) luminosity distributions of quiescent red sequence galaxies (red) and active blue sequence galaxies (blue). Also shown is the luminosity distribution of red sequence galaxies with $\log$($L_{24}$/$L_{3.6}$)$>$-0.7 classified as either star-forming (orange) or AGN (cyan) based on the spectral line analysis.}
\end{center}
\end{figure}

Since we are measuring extinction values based on emission lines, we are in effect applying an extinction correction based on the star-forming regions of a galaxy rather than the average extinction of the galaxy as a whole. If the red optical colours of star-forming galaxies are due to a bulge component, while intense bursts of star-formation activity are due to a disk component, such a disk component may lead to over corrections for extinction, which may disguise the fact that the red optical colour of the galaxy is due to a bulge contribution (i.e. old stellar populations).

\subsection{Comparison of Optical and Infrared Luminosities}

Red sequence star-forming galaxies with $\log$($L_{24}$/$L_{3.6}$)$>$-0.7 are found to have similar optical colours as early-type galaxies, whereas their infrared colours most resemble that of late-type galaxies. We therefore compare the optical and infrared spectral energy distributions of these systems with that of galaxies in the red and blue sequence as a whole, in terms of rest-frame optical and infrared luminosities (see $\S$3).

Figures 7a and 7b show the optical and infrared luminosity distributions of our sample.
Red sequence galaxies with quiescent star-formation activity (red) are generally found to have higher optical luminosities than blue sequence star-forming galaxies. 
Star-forming galaxies with red optical colours (orange) and $\log$($L_{24}$/$L_{3.6}$)$>$-0.7 are not only found to have a similar optical luminosity distribution as bulge-dominated quiescent red sequence galaxies, but optical luminosities that are higher than the majority of star-forming galaxies in the blue sequence.

The infrared luminosity distribution of red sequence galaxies is on average found to be lower than star-forming galaxies in the blue sequence, since much of the infrared activity in the red sequence is due to quiescent star-formation. However, approximately a quarter of red sequence galaxies are found to have infrared luminosities higher than the majority of star-forming galaxies in the blue sequence. These galaxies either have an AGN contribution, where dust heated by the AGN accounts for their high infrared luminosities, or these galaxies are undergoing intense star-formation activity.

\subsection{SWIRE$_{-}$J104152.92+595616.3}

Figure 8 shows an example of a red sequence star-forming galaxy at $z$ $\sim$ 0.147 with $\log$($L_{24}$/$L_{3.6}$)$>$-0.7. This galaxy has a S$\acute{{e}}$rsic index $\sim$ 3.6 (see $\S$6) and the SDSS and 3.6$\mu$m SWIRE images (Figure 8a) show a bulge-component which would account for its red optical colour. However, this galaxy is also found to have strong mid-to-far infrared emission as illustrated at 70$\mu$m (\textit{f$_{70\mu}$$_{m}$} $\sim$ 30mJy). We model the optical and infrared spectral energy distribution of this galaxy using the SWIRE template library (Figure 8b).  We have already shown using spectral line diagnostics (see $\S$4) that these galaxies are not contaminated by AGN. Therefore, as a test, we attempt to model the infrared SED of this galaxy as Seyfert 2 AGN (green), and find that doing so would significantly overestimate its far-infrared emission. We find the infrared SED is best modelled as a Spiral - Sc (blue), which would agree with the fact that these systems have high infrared emission as a result of bursts of star-formation activity.

\begin{figure}
\begin{center}
\includegraphics[height=3.8cm,width=9cm]{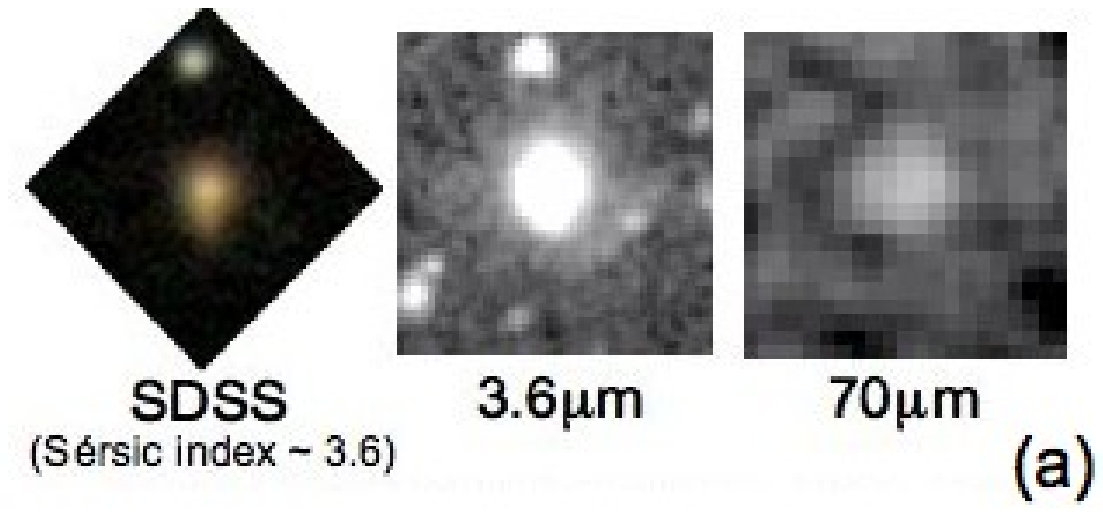} 
\includegraphics[height=7cm,width=9cm]{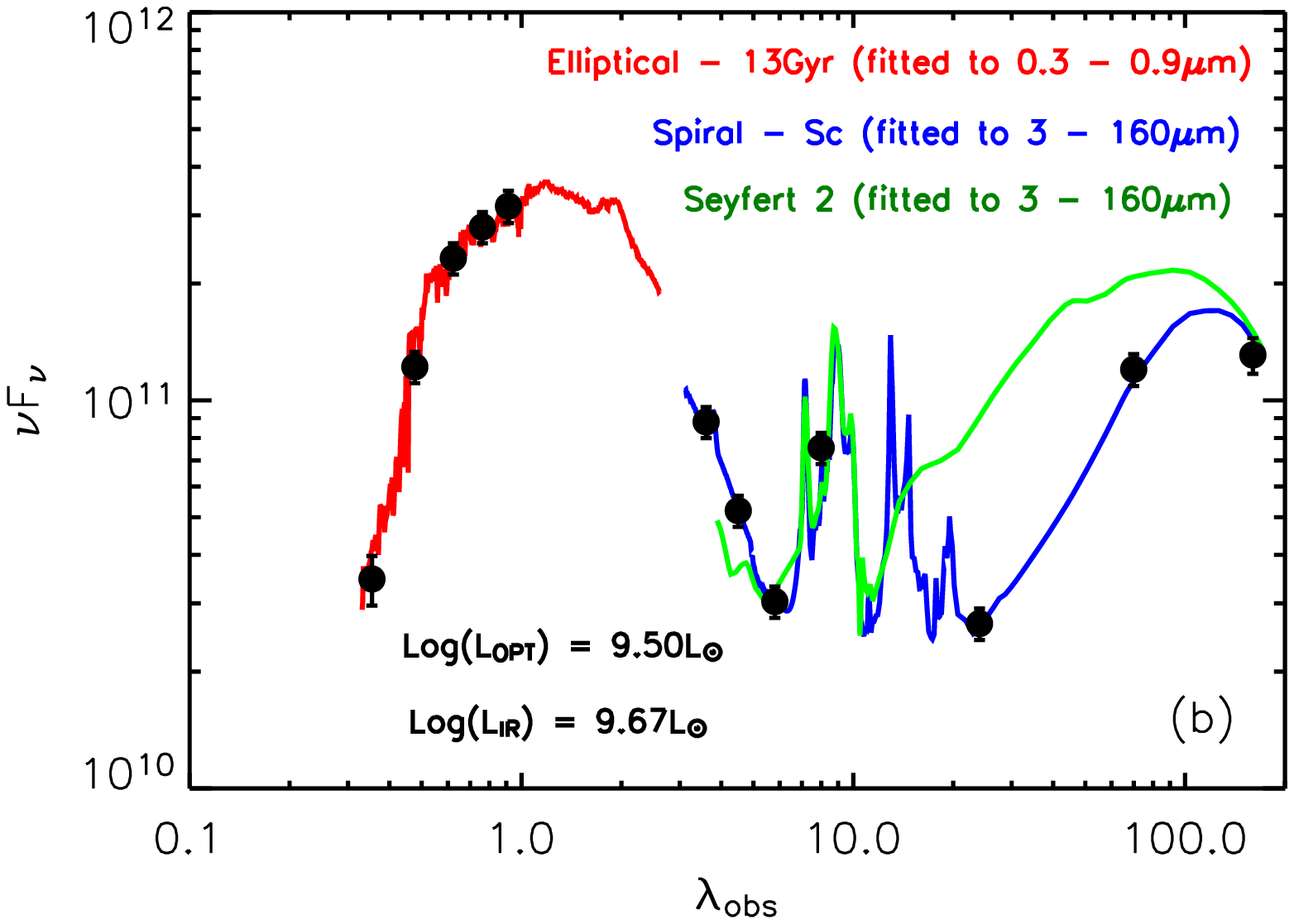}
\caption{SWIRE$_{-}$J104152.92+595616.3 - Example of red sequence star-forming galaxy with $\log$($L_{24}$/$L_{3.6}$)$>$-0.7. (a) SDSS optical image, 3.6$\mu$m and 70$\mu$m SWIRE images. (b) The optical and infrared spectral energy distribution modelled with SWIRE galaxy templates.}
\end{center}
\end{figure}

The optical SED of this galaxy is found to resemble that of a 13Gyr Elliptical template (red), and the infrared SED best modelled by a Spiral template (blue). This galaxy did not meet the 3$\sigma$ emission line criteria (see $\S7.1$) required for determining an estimate of extinction using the Balmer Decrement. As mentioned in $\S$7.1, this could either be due to the fact that the galaxy is so extinct that its Balmer lines have been suppressed, or that the galaxy is characterised by weak Balmer emission. Based on the high optical luminosity of this system, a surface brightness profile resembling an elliptical galaxy (i.e. a S$\acute{{e}}$rsic index $>$ 1.8) and an optical SED best modelled as an elliptical, these are indications that the red optical colours of such systems are mainly a result of a bulge-component rather than substantial reddening.

\begin{table*}
\caption{Number density and luminosity density of red sequence and blue sequence galaxies. Number and luminosity densities as a percentage of the sample total are given in brackets.}
\begin{center}
\begin{tabular}{c c c c c c}
\hline\hline
SEQUENCE & $\log$($L_{24}$/$L_{3.6}$)$<$-0.7 & $\log$($L_{24}$/$L_{3.6}$)$>$-0.7 & $\log$($L_{24}$/$L_{3.6}$)$>$-0.7 & SEQUENCE & SAMPLE\\
&  & (AGN) & (Star-forming) & TOTAL & TOTAL\\ 
& & & & (h$^{3}$ Mpc$^{-3}$) & (h$^{3}$ Mpc$^{-3}$) \\
\\
 \cline{2-5}
 \\
& & \multicolumn{2}{c}{NUMBER DENSITY}  & &\\
 \\
 \cline{2-5}
 \\ 
RED & 69$\%$ (55$\%$) & 14$\%$ (11$\%$) & 17$\%$ (13$\%$) & 3.5x10$^{-6}$ & (4.4x10$^{-6}$)\\
BLUE & 10$\%$ (2$\%$) & 7$\%$ (1$\%$) & 83$\%$ (18$\%$) & 8.9x10$^{-7}$ &\\
\\
 \cline{1-6}
 \\ 
& & \multicolumn{2}{c}{OPTICAL LUMINOSITY DENSITY} & &\\
\\
 \cline{2-5}
 \\ 
RED & 69$\%$ (67$\%$) & 15$\%$ (15$\%$) & 16$\%$ (16$\%$) & 5.6x10$^{3}$$L_{\sun}$ & (6.8x10$^{3}$$L_{\sun}$)\\
BLUE & 13$\%$ ($<$1$\%$) & 6$\%$ ($<$1$\%$) & 81$\%$ (2$\%$) & 1.2x10$^{3}$$L_{\sun}$ &\\
\\
 \cline{1-6}
 \\ 
& & \multicolumn{2}{c}{INFRARED LUMINOSITY DENSITY} & &\\
\\
 \cline{2-5}
 \\ 
RED & 27$\%$ (17$\%$) & 29$\%$ (19$\%$) & 44$\%$ (28$\%$) & 3.8x10$^{3}$$L_{\sun}$ & (6.1x10$^{3}$$L_{\sun}$) \\
BLUE & 1$\%$ ($<$1$\%$) & 6$\%$ (2$\%$) & 93$\%$ (34$\%$) & 2.3x10$^{3}$$L_{\sun}$ &\\

\hline
\end{tabular}
\end{center}
\end{table*}

\section{The Number Density and Luminosity Density of red and blue sequence galaxies}

Star-forming galaxies are expected to provide a major contribution to both the number density and optical/infrared luminosity densities of blue sequence galaxies. In comparison, the majority of red sequence galaxies undergoing quiescent star-formation activity will account for the number density and optical luminosity density of the red sequence, providing less of a contribution to the infrared luminosity density than more active systems. However, infrared colours have shown that a quarter of red sequence galaxies in our sample are active systems, and must therefore provide a significant contribution to both the number density and luminosity density of the local universe.

We therefore determine the integrated number density - total of 1/$V_{\rm max}$ (Schmidt 1968; Felten 1976), of galaxies in the red and blue sequence and compare with the number density of AGN and star-forming galaxies with red optical colours and $\log$($L_{24}$/$L_{3.6}$) $>$ -0.7. $V_{\rm max_{i}}$ is the volume corresponding to the maximum redshift at which a source could be detected by the survey. We set this maximum redshift by considering optical and infrared limits, determined using k-corrections calculated for each galaxy from its optical and infrared template fits (see $\S$3).

For each galaxy, the density contribution 1/$V_{\rm max_{i}}$ and the luminosity contribution $L_{i}$/$V_{\rm max_{i}}$ are computed. We correct for incompleteness as a result of the spectroscopic cuts in ZWARNING and zConf (see $\S$2.2).

Table 4 shows the number and luminosity densities of red and blue sequence galaxies in our sample. 
Galaxies with red optical colours and $\log$($L_{24}$/$L_{3.6}$) $<$ -0.7 account for more than two-thirds of the number density and optical luminosity density of red sequence galaxies. These sources make up more than half the number density and more than two-thirds of the optical luminosity density of our total galaxy sample. In contrast, these galaxies account for only 17$\%$ of the total infrared luminosity density due to their quiescent star-formation activity.

Whereas the number density and optical luminosity density of our galaxy sample is dominated by red sequence galaxies with quiescent star-formaton, the contribution to the infrared luminosity density comes from a number of populations in both the red and blue sequence. Star-forming galaxies account for more than 90$\%$ of the infrared luminosity density of the blue sequence, and 34$\%$ of the infrared luminosity density of our total galaxy sample. These systems account for only 2$\%$ of the total optical luminosity density. 

We find that active systems (AGN and star-forming galaxies) in the red sequence account for a quarter of the total number density of our galaxy sample; AGN representing 11$\%$ and star-forming galaxies 13$\%$. 

Thermal dust emission from red sequence AGN contributes $\sim$19$\%$ to the total infrared luminosity density. Due to intense bursts of star-formation activity, red sequence star-forming galaxies contribute $\sim$28$\%$ to the infrared luminosity density of our SWIRE/SDSS sample. In addition, the optical luminosity density of these star-forming systems are found to be eight times greater than that of star-forming galaxies in the blue sequence.

\section{Discussion and Conclusions}

We present the analysis of the rest-frame optical and infrared colours of 1114 $z$$<$0.3 galaxies from the Spitzer Wide-Area InfraRed Extragalactic survey (SWIRE) and the Sloan Digital Sky Survey (SDSS). 

We separate our sample into two populations. Based on rest-frame (g-r) colour, galaxies with red optical colours are defined as `red sequence galaxies'. Galaxies with blue optical colours are defined as `blue sequence galaxies'.

To test whether the bi-modal sequence seen using rest-frame optical colours represent a simple division of early and late-type galaxies in the local universe, we analyse their optical spectra to identify quiescent and active (starburst and AGN) systems and use infrared $\log$($L_{24}$/$L_{3.6}$) colour to investigate the star-formation activity of each sequence as a function of stellar mass. 

We find red sequence galaxies have a broad range of star-formation activity independent of their stellar mass. Galaxies with weak star-formation appear to be bulge-dominated early-type systems, whereas those with higher levels of activity have a contribution from both a bulge and disk component.
In comparison, galaxies in the blue sequence are found to have high levels of activity, resembling disk-dominated spiral-like systems. 

We therefore postulate that the (g-r) colour and $\log$($L_{24}$/$L_{3.6}$) colour of galaxies in our sample are determined primarily by a bulge-to-disk ratio. This relationship is perhaps more continuous than is perceived when viewed in the projection of one colour, and is such that the (g-r) colour is sensitive to the bulge-to-disk ratio for disk-dominated galaxies, whereas the $\log$($L_{24}$/$L_{3.6}$) colour is sensitive for bulge-dominated systems.

We find a significant fraction of our red sequence sample ($\sim$26$\%$, 195 sources) with intrinsic red optical colours and high $\log$($L_{24}$/$L_{3.6}$) ratio. These systems are characterised as either AGN (78 sources) or systems with bursts of star-formation activity (117 sources). AGN owe their red optical colours to their bulge component, and their high $\log$($L_{24}$/$L_{3.6}$) ratio would be due to dust heated by AGN.  Using galaxy template models, star-forming galaxies in the red sequence are found to have up to a 90$\%$ Elliptical-like contribution and a 10$\%$ starburst contribution. The red optical colours of these systems would therefore be due to a bulge component, and their high $\log$($L_{24}$/$L_{3.6}$) ratio would correspond to intense star-formation activity accounted for by a disk component. 

Extinction could in principle account for the red optical colours of approximately half of these star-forming galaxies. However, such corrections for extinction are based on the star-forming regions of a galaxy rather than the average extinction of the galaxy as a whole. If these systems are undergoing localised bursts of star-formation activity, this would lead to over-corrections for extinction, which may disguise the fact that the red optical colours of these star-forming galaxies are actually due to a significant bulge component. 

To gain a better understanding of the nature of these red sequence star-forming galaxies, we compare their optical and infrared luminosities with that of red and blue sequence galaxies as a whole.
These systems are found to have optical luminosities resembling that of bulge dominated red sequence galaxies with quiescent star-formation, whereas intense bursts of star-formation activity means their infrared luminosities are on average higher than the majority of star-forming galaxies in the blue sequence. This confirms our suspicion that these galaxies have significant bulge components and are not disk galaxies with high levels of extinction.

It has been postulated that the recent evolution of early-type galaxy populations can be explained by a `dry-mergers scenario' (see e.g. Toomre \& Toomre 1972; Bell et al.\ 2006), where non-star-forming galaxies merge to form massive early-type systems. 
Early studies of galaxy morphology and density in rich clusters (Dressler 1980, Postman \& Geller 1984) have also shown that the large bulge-to-disk ratio commonly associated with early-type galaxies is inconsistent with the idea that such systems are the progenitors of disk dominated late-type galaxies. More recent studies have shown that star-forming spiral-spiral mergers could result in the formation of early-type galaxies at low optical luminosities (Naab et al.\ 2006). However, semi-analytic work by Khochfar \& Burkert (2003, 2005) and optical survey studies by van Dokkum et al.\ (1999); Tran et al.\ (2005) and Bell et al.\ (2005) have suggested that there are relatively few optically blue star-forming galaxies in the low redshift universe luminous enough to merge into present day massive early-type galaxies, favouring the dry-mergers scenario.  Our results suggest that there is a substantial population of massive, red, galaxies with ongoing star-formation activity. This may reduce the need for dry-mergers.

We finally quantify the contribution of these dusty star-forming galaxies to optically selected galaxies in the local universe. Red sequence galaxies make up 79$\%$ of the total number density of our sample. 83$\%$ of blue sequence galaxies and 17$\%$ of red sequence galaxies are found to be actively star-forming systems. Most of the optical luminosity density (67$\%$) is due to quiescent red sequence galaxies and less than 4$\%$ to galaxies in the blue sequence. Star-forming galaxies are found to have a similar contribution to the optical luminosity density of the red sequence as AGN. Red sequence AGN and quiescent galaxies are responsible for 36$\%$ of the total infrared luminosity density. Active star-forming galaxies produce 62$\%$ of the total infrared luminosity density of our SWIRE/SDSS sample, of which 28$\%$ is as a result of star-forming galaxies in the red sequence.

\section*{Acknowledgments}

PD is supported by PPARC Studentship (PPA/S/S/2002/03500/), SJO is
supported by a Leverhulme Research Fellowship, SJO and IW are
supported by PPARC standard grant (PPA/G/S/2000/00508 \& PPA/G/S/2002/00481). We thank 
Jon Loveday for his advice on certain aspects of the SDSS data. 
We thank the referee Randall Rojas for his helpful comments and suggestions.
Support for this work, part of the Spitzer Space Telescope Legacy
Science Program, was provided by NASA through an award issued by the
Jet Propulsion Laboratory, California Institute of Technology under
NASA contract 1407. 

\bibliographystyle{mn2e}

\label{lastpage}

\end{document}